\documentclass{article}

\usepackage{arxiv}

\usepackage[utf8]{inputenc} 
\usepackage[T1]{fontenc}    
\usepackage[hypertexnames=false]{hyperref}       
\usepackage{url}            
\usepackage{booktabs}       
\usepackage{amsfonts}       
\usepackage{nicefrac}       
\usepackage{microtype}      
\usepackage{graphicx}
\usepackage{natbib}
\usepackage{doi}
\usepackage{array}
\usepackage{multirow}
\usepackage{mathtools}
\usepackage{amsmath}
\usepackage{adjustbox}
\usepackage{afterpage}
\usepackage{icomma}
\usepackage{nicematrix}
\usepackage{xltabular}
\usepackage{caption}

\title{Uncovering All Highly Credible Binary Treatment Hierarchy Questions in Network Meta-Analysis}

\author{ \href{https://orcid.org/0000-0002-8272-797X}{\includegraphics[scale=0.06]{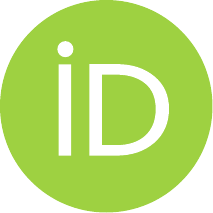}\hspace{1mm}Caitlin H. Daly} \\
	Department of Statistics and Actuarial Science\\
	University of Waterloo\\
	Waterloo, Ontario\\
    N2L 3G1\\
    Canada\\
    \And
    Chloe Tan \\
	Department of Statistics and Actuarial Science\\
	University of Waterloo\\
	Waterloo, Ontario\\
    N2L 3G1\\
    Canada\\
	\And
	\href{https://orcid.org/0000-0003-4124-2498}{\includegraphics[scale=0.06]{images/orcid.pdf}\hspace{1mm}Audrey B{\'e}liveau} \thanks{Corresponding Author: \texttt{audrey.beliveau@uwaterloo.ca}} \\
	Department of Statistics and Actuarial Science\\
	University of Waterloo\\
	Waterloo, Ontario\\
    N2L 3G1\\
    Canada\\
}

\date{}

\renewcommand{\headeright}{}
\renewcommand{\undertitle}{}
\renewcommand{\shorttitle}{Binary Treatment Hierarchy Questions in NMA}

\newcommand{\ep}{\hat\pi}
\newcommand{\prob}{\pi}

\hypersetup{
pdftitle={Uncovering All Highly Credible Binary Treatment Hierarchy Questions in Network Meta-Analysis},
pdfauthor={Caitlin H.~Daly, Chloe Tan, Audrey B{\'e}liveau},
pdfkeywords={combinatorial explosion; network meta-analysis; posterior probability; ranking; search methods; treatment hierarchy questions},
}

\begin{document}
\maketitle

\begin{abstract}
In recent years, there has been growing research interest in addressing treatment hierarchy questions within network meta-analysis (NMA). In NMAs involving many treatments, the number of possible hierarchy questions becomes prohibitively large. To manage this complexity, previous work has recommended pre-selecting specific hierarchy questions of interest (e.g., ``among options A, B, C, D, E, do treatments A and B have the two best effects in terms of improving outcome X?") and calculating the empirical probabilities of the answers being true given the data. In contrast, we propose an efficient and scalable algorithmic approach that eliminates the need for pre-specification by systematically generating a comprehensive catalog of highly credible treatment hierarchy questions, specifically, those with empirical probabilities exceeding a chosen threshold (e.g., 95\%). This enables decision-makers to extract all meaningful insights supported by the data. An additional algorithm trims redundant insights from the output to facilitate interpretation. We define and address six broad types of binary hierarchy questions (i.e., those with true/false answers), covering standard hierarchy questions answered using existing ranking metrics - pairwise comparisons and (cumulative) ranking probabilities - as well as many other complex hierarchy questions. We have implemented our methods in an R package and illustrate their application using real NMA datasets on diabetes and depression interventions. Beyond NMA, our approach is relevant to any decision problem concerning three or more treatment options.
\end{abstract}

\keywords{combinatorial explosion \and network meta-analysis \and posterior probability \and ranking \and search methods \and treatment hierarchy questions}

\section{Introduction} \label{s:Intro}

A rational decision involves formulating a hierarchical set of options that reflects the preferred order of each option's expected consequences, then selecting the optimal option \citep{march}. In medicine, there are often multiple competing options to treat a health concern. A treatment hierarchy reflects the rank-ordered performance of a set of treatment options from best to worst, where performance may be defined based on a single outcome quantifying efficacy, safety, or tolerability, or a combination of several outcomes, and performances are distinguished based on some criterion (e.g., a numerical comparison reflecting a minimally important difference [MID]) \citep{papakonstantinou2022}. For the same set of treatment options, hierarchies will vary across populations, interventional settings, outcomes, and outcome measures.

Even under a single decision problem defined by a specific population, set of treatment options, outcome, measurement tool, and other circumstances, several questions regarding features of a true treatment hierarchy may be asked. Those familiar with combinatorics are well aware there are multiple ways of arranging and selecting from a set of options. Treatment hierarchy questions may focus on a permutation of the full set of options, e.g., ``among options A, B, C, D, E, do their effects in terms of improving outcome X suggest A~$>$~B~$>$~C~$>$~D~$>$~E?", or a broader question regarding a specific class of treatments, e.g., ``among options A, B, C, D, E, do treatments A, B, and D have the three best effects in terms of improving outcome X?".

To answer these questions, evidence on the comparative effectiveness of all treatment options is required. Network meta-analysis (NMA) may be used to synthesize such evidence from multiple sources; it enables the simultaneous comparison of multiple treatments that may or may not have been directly compared in head-to-head clinical trials \citep{lu2004, caldwell2005}. In recent years, there has been significant research interest in exploring treatment hierarchy questions within NMA. Salanti et al. (2022) first formalized the concept of a treatment hierarchy question in NMA to link it to a meaningful clinical research question, arguing it should explicitly define the endpoint (or outcome), how it is summarized for each treatment, the effect measure used to contrast outcomes between treatments, and the criterion used to prefer one treatment over another (i.e., ranking metric) \citep{salanti2022_hierachyq}. Papakonstantinou et al. (2022) expanded on this to answer questions consisting of a criterion beyond the ranking metrics available in NMA \citep{papakonstantinou2022}. Their approach acknowledges the existence of a true underlying treatment hierarchy, which may be wholly or partly of interest based on some pre-specified research question(s) of interest; the empirical probabilities of all hierarchies meeting a question's criterion are calculated to assess its certainty. 

Unless there is strong evidence revealing plausible treatment hierarchies a priori, decision makers may approach evidence with an open mind, with no firm preconceived notions of what the true underlying treatment hierarchy may be. As such, it may be difficult to pre-specify treatment hierarchies of interest. In addition, the full spectrum of hierarchy questions that may be answered by NMA has yet to be defined \citep{salanti2022_hierachyq}. Our goal is to identify and formalize a broad class of treatment hierarchy questions that are of interest to NMA, and to develop a computationally efficient method for compiling a comprehensive catalog of questions from this class that are strongly supported by the data, i.e. highly credible, for instance with empirical probabilities exceeding 95\%. While our methods are developed in the context of NMA, the underlying principles are broadly applicable to any setting involving comparisons among multiple treatments such as regression with categorical covariates.

We focus specifically on what we term ``binary" treatment hierarchy questions, which are true-or-false questions about characteristics of the true underlying treatment hierarchy (as opposed to data-driven summaries, such as whether a treatment’s posterior mean rank falls within a certain range). Our approach was developed with a Bayesian framework in mind, and hence we primarily use Bayesian terminology in this paper. Nevertheless, our approach is also applicable to samples from an assumed joint probability distribution of the relative treatment effects estimated in a frequentist NMA. Through posterior probabilities, the answers to binary treatment hierarchy questions provide an objective means of assessing the credibility of what the evidence appears to reveal - this may be helpful when multiple decision makers (e.g., guideline committees) with different experiences and tolerances of uncertainty work through their conflicting subjective interpretations to come to a consensus on treatment recommendations based on the same set of evidence \citep{small2014}. Although these questions can be viewed as statistical hypotheses, our approach is not framed as hypothesis testing. No hypotheses are pre-specified. Instead, it serves as a descriptive summary of the evidence, such as the joint posterior distribution of the relative treatment effects derived from a Bayesian NMA, providing consideration to all possible hierarchy questions. Issues related to multiple testing adjustments are acknowledged but left for future exploration. 

Section \ref{s:binQuest} introduces a taxonomy for binary treatment hierarchy questions, introduces relevant notations, and quantifies how many unique questions fall within each category. Section \ref{s:algo_descriptions} introduces three computational strategies for identifying which of these questions are strongly supported by evidence from an NMA, as well as strategies to trim redundant credible questions among those identified. Section \ref{s:empIll} applies these methods to an NMA dataset and reports the findings. Finally, a discussion ensues in Section \ref{s:discuss}.

\section{Taxonomy and Abundance of Binary Treatment Hierarchy Questions} \label{s:binQuest}

We assume the decision problem has been defined in terms of a specific target population, outcome, and other settings for which an underlying true treatment hierarchy is of interest. In this section, we define six classes of binary treatment hierarchy questions in light of this decision problem. In Section \ref{sm:1_reporttools} of the Supplementary Materials, we describe how the binary treatment hierarchy questions defined in Sections \ref{s:rankPerm} - \ref{s:rankSets} align with established and commonly reported NMA summaries. Questions that are not covered in this work are discussed in Section \ref{s:discuss}. Additionally, we propose unified notations for referencing binary treatment hierarchy questions and we determine the number of meaningful and unique questions within each type, accounting for clear cases where questions involving a subset of treatments are equivalent to questions involving a different subset of treatments. We present a more thorough account of redundant (but not necessarily equivalent) questions in Sections \ref{s:trim} and \ref{sm:2_trim} of the Supplementary Materials.

In all cases, we assume it is known whether smaller or larger outcomes are preferable, and that the MID \citep{jayadevappa2017}, if relevant, is specified, otherwise we assume MID = 0. In addition, we assume rank 1 corresponds to a treatment with the best performance, while rank $n$ corresponds to a treatment with the worst performance, where $n$ is the number of treatments under consideration within set $\mathcal{T}$ (e.g., $\mathcal{T} = \{$A, B, C, D, E$\}$ is a set of $n=5$ treatments).

\subsection{Ranked Permutations of Treatments} \label{s:rankPerm} \par

A treatment hierarchy question of interest might ask: ``Do treatments A, B, and D have ranks 1, 2, and 3, respectively?" Generally, questions of this type examine whether treatments in a permutation (e.g. (A, B, D)) hold specific consecutive ranks in a directed order (e.g. 1, 2 and 3 for A, B and D, respectively). For simplicity, the question above is succinctly represented as $(\text{A, B, D})_1^3$. More generally, we refer to such questions as ``ranked permutations" and denote them individually as $\mathcal P_i^j$, where $\mathcal P$ represents a permutation of at least 2 treatments ranked from $i$ to $j$, with $1 \leq i < j \leq n$.

Considering all possible values for $\mathcal P$, $i$, and $j$, the total unique number of treatment hierarchy questions concerning ranked permutations is given by $n(n-1)^2 + n(n-1)(n-2)^2 + ... + n(n-1)\cdots3^2 + n!$. All terms apart from the last of this expression are derived by first calculating the number of permutations of size 2 through \( n-2 \), which are \( \frac{n!}{n-2!}, \frac{n!}{n-3!}, \dots, \frac{n!}{2!} \), respectively. Each of these is then multiplied by the corresponding number of ranking ranges: \( n-1 \) for size 2, \( n-2 \) for size 3, and so on, down to 3 for size \( n-2 \). The last term, $n!$, accounts for ranked permutations of size $n$, which answers questions equivalent to ranked permutations of size $n-1$ (e.g., if $\mathcal{T} = \{$A, B, C, D, E$\}$, (A, B, C, D, E)$_1^5 \iff$ (A, B, C, D)$_1^4 \iff$ (B, C, D, E)$_2^5$). 

\subsection{Permutations of Treatments}\label{s:perm} \par

Another treatment hierarchy question of interest could be of the type: ``Do treatments A, B, and D rank consecutively and in this specific order (from best to worst)?" This question differs from the question in Section \ref{s:rankPerm} in that specific ranks may be overlooked; for example, a decision maker may be interested in whether treatments belonging to a drug class are ranked similarly or together and in an expected order. For simplicity, we denote this question as (A, B, D), and for a general permutation $\mathcal P$ of treatments, with $\text{size}(\mathcal P) = 2,~\dots,~ n-1$, we refer to the equivalent treatment hierarchy question as $\mathcal P$. Considering all possible sizes of $\mathcal P$, the total of such treatment hierarchy questions is given by the sum $\frac{n!}{(n-2)!} + \frac{n!}{(n-3)!} + ... + \frac{n!}{2!} + n!$. Permutations of size $n$ answer equivalent questions as ranked permutations of size $n$ and are not included in this count, as they are accounted for in Section~\ref{s:rankPerm}.

\subsection{Partial Hierarchies of Treatments}\label{s:poset} \par

A third treatment hierarchy question of interest is similar to the previous one, but does not require the treatments to be consecutive. This question is of the type: ``Do treatments A, B, and D rank in this order amongst themselves, from better to worse?" Such questions may be posed when a subset of the treatments is of interest (but other treatments are required to connect the network), and so only the ordering within the subset is of interest. We denote the question A $>$ B $>$ D as $h((\text{A,~B,~D}), 0)$, where $h$ is a function constructing the partial hierarchy from the permutation of treatments $(\text{A,~B,~D})$ and the MID used to qualify a treatment difference, which is 0 in the example question since it is unspecified. For a general permutation $\mathcal P$ of treatments from $\mathcal T$ of size($\mathcal P) = 2, ~\dots,~ n-1$, we can denote the partial hierarchy question as $h(\mathcal P, \text{MID})$. Note we do not include partial hierarchies of size $n$ in this definition as these answer equivalent questions as ranked permutations of size $n$. 

Considering all possible sizes of $\mathcal P$ at a fixed MID, there are $\binom{n}{2} + \frac{n!}{(n-3)!} + ... + n!$ partial hierarchy questions, as we sum all permutations, except for size 2 where by symmetry we only need to investigate half of them to know the other half via the complement.    

\subsection{Ranked Combinations of Treatments}\label{s:rankComb} \par

Another question we could ask is: ``Do treatments A, B, and D rank in any order amongst the ranks 1, 2, and 3?". In other words, ``are treatments A, B, and D among the top 3"? We denote this question as $\{\text{A, B, D}\}_1^3$, where $\{\text{A, B, D}\}$ is a combination and the subscript and superscript represent the ranking range. In the general case, for any combination $\mathcal C$ of treatments, with size$(\mathcal C) \in \{2, 3, ..., n-1\}$ and ranking range $1 \leq i < j \leq n$, we denote the treatment hierarchy question as $\mathcal C_i^j$. Note ranked combinations of size $n$ are excluded as these do not answer relevant questions (i.e., do all treatments rank 1 through $n$ in any order?). 

Considering all possible values for $\mathcal C$, $i$ and $j$, there are $(n-2)\binom{n}{2} + (n-3)\binom{n}{3} + ... + 2\binom{n}{n-2} +2n = n(2^{n-1}-n+1)$ such treatment hierarchy questions. This follows the same reasoning as ranked permutations in Section \ref{s:rankPerm}, but instead of permutations, we sum the number of combinations of each size modulated by their respective number of ranking ranges. However, some of these questions are redundant because top-ranked combinations $\mathcal C_1^{j}$ and bottom-ranked combinations $\{\mathcal{C'}\}_{j+1}^n$ are effectively the same question, for $j=2$ to $n-2$, where $\mathcal{C'} = \mathcal{T}\backslash\mathcal{C}$.

\subsection{Combinations of Treatments}\label{s:comb} \par

Another treatment hierarchy question of interest could be of the type: ``Do treatments A, B, and D rank consecutively (but in any order)?" We denote this question as $\{\text{A, B, D}\}$, where $\{\text{A, B, D}\}$ is a combination. More generally, for any combination $\mathcal C$ of treatments, with size$(\mathcal C) \in \{2, 3, ..., n-1\}$, we denote the treatment hierarchy question as $\mathcal C$. Considering all possible sizes of $\mathcal C$, we sum them all to get $\binom{n}{2} + \binom{n}{3} + ... + \binom{n}{n-1} = 2^{n} - n - 2$ treatment hierarchy questions. Note there is only one way to have a combination of size $n$ and this is excluded as it does not answer a relevant question (i.e., do all treatments rank consecutively in any order?).

\subsection{Sets of Ranks for Individual Treatments}\label{s:rankSets} \par

All treatment hierarchy questions defined in Sections \ref{s:rankPerm} to \ref{s:comb} pertain to at least two treatments. Now, as a final question type, we examine whether an individual treatment’s rank belongs to a specified set. For instance, we might ask, “Is treatment A the best treatment?” or “Is treatment B among the top 3 treatments?” We denote such treatment hierarchy questions by \( T_\mathcal{R} \), where $T \in \mathcal{T}$ represents an individual treatment and $\mathcal{R} \subset \{1, 2, ..., n\}$ is a proper non-empty subset of ranks. Under this notation, the two examples above can be expressed as \( \text{A}_{\{1\}} \) and \( \text{B}_{\{1,2,3\}} \). For each of the \( n \) treatments, there are \( 2^n -2\) proper non-empty subsets of ranks, resulting in a total of \( n(2^n - 2) \) relevant treatment hierarchy questions pertaining to sets of ranks for individual treatments. 

Some of these $T_{\mathcal R}$ questions may initially appear uninteresting, for instance, $T_\mathcal R = \text{A}_{\{1,3,5,7\}}$, i.e. “Does treatment A have an odd numbered rank?”. However, we do not wish to limit $\mathcal R$ to be a consecutive set of ranks because ranking distributions may be multimodal~\citep{wu2021_entropy}, particularly when precision varies across trials. Such $T_\mathcal R$ questions, if they show high empirical probability, would be relevant to examine. Since ranking distributions are not known a priori, we cannot justify excluding any $T_\mathcal R$ questions a priori. However, in Section \ref{s:algo3}, we will propose a procedure to extract only the most meaningful among all $T_{\mathcal R}$ hierarchy questions, at a given credibility threshold.

\section{Methods for Uncovering Highly Credible Hierarchy Questions} \label{s:algo_descriptions}

\begin{figure}
	\begin{center}
	\includegraphics[scale = 1]{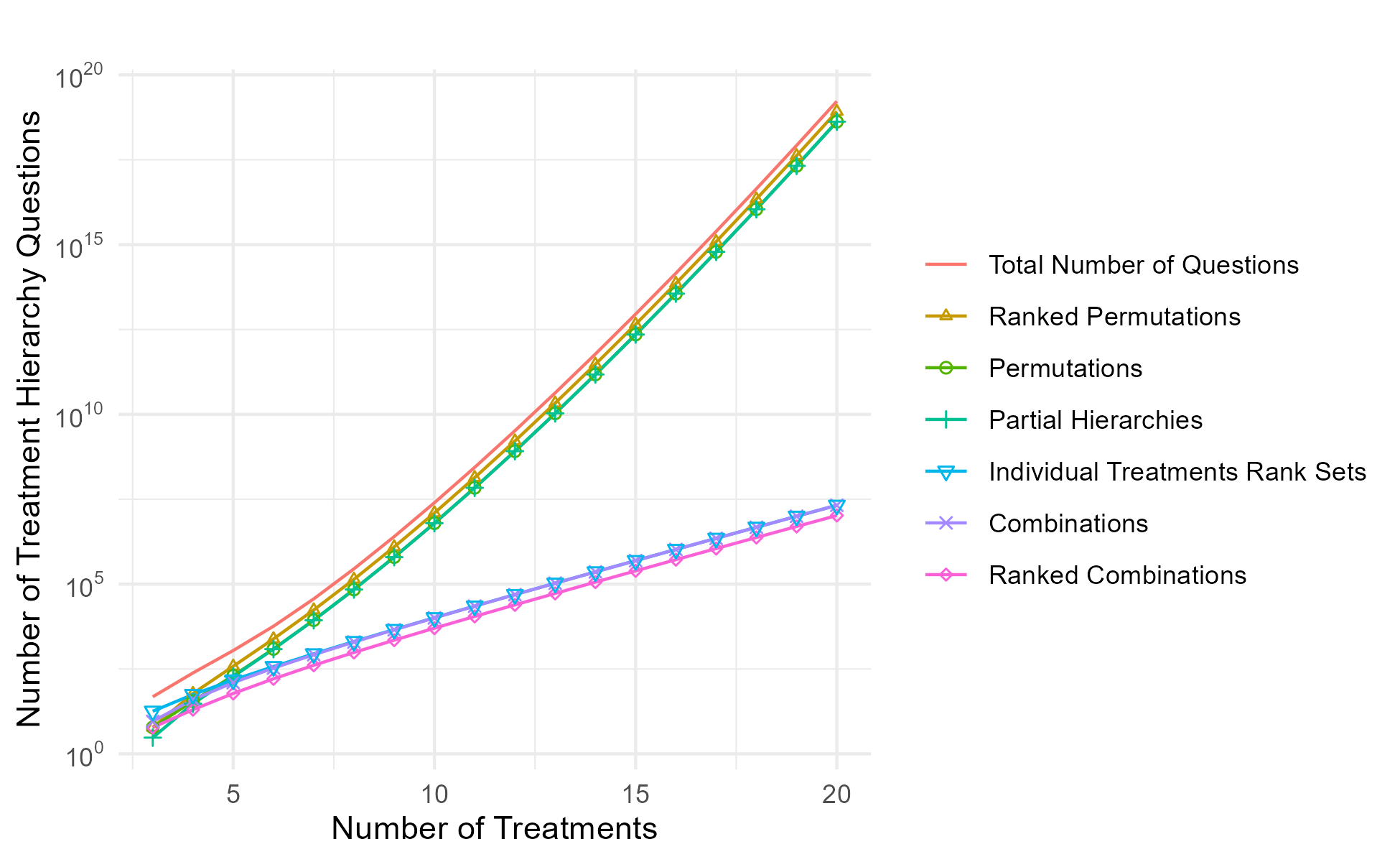}
	\caption{The number of binary treatment hierarchy questions to evaluate via brute force, expressed as a power of 10, in relation to the number of treatments in the NMA.}
	\label{fig:num_of_questions}
    \end{center}
\end{figure}

In Section \ref{s:binQuest}, we identified six broad types of binary treatment hierarchy questions within an NMA. Figure \ref{fig:num_of_questions} demonstrates the computational challenges involved in evaluating the empirical probabilities for all such questions. For example, the total number of treatment hierarchy questions exceeds one million in networks of size $n=9$, and further increases drastically as $n$ increases. To systematically calculate the empirical probability of each question being true via brute force, the time complexity would be $O(n!K)$ for ranked permutation, permutation and partial hierarchy questions, $O(n2^{n-1}K)$ for ranked combination questions, $O(2^nK)$ for combination questions and $O(n2^nK)$ for individual treatment rankings. Here, $K$ represents the number of samples used to calculate these probabilities, typically obtained via Markov Chain Monte Carlo (MCMC) chains of the joint posterior distribution of relative treatment effects in Bayesian NMA, which we focus on in this paper. Alternatively, these may be drawn independently from the joint probability density of relative treatment effect estimates in a frequentist NMA. The degree of complexity becomes computationally impractical for networks with a moderate to large number of treatments, as it increases rapidly with $n$.

Let $\mathcal Q$ denote a binary treatment hierarchy question of interest and define the indicator function $\delta(\mathcal Q) = 1$ if the answer to $\mathcal Q$ is true under the decision problem of the NMA, and 0 otherwise. We are interested in the posterior probability $\prob(\delta(\mathcal Q) = 1 | \text{data})$. Following \citet{papakonstantinou2022}, an estimate is obtained from $K$ posterior samples by calculating the empirical probability $\ep(\mathcal Q) = \frac{1}{K} \sum_{k=1}^{K}I_k(\mathcal Q)$, where $I_k(\mathcal Q)$ is an indicator that question $\mathcal Q$ is true in posterior sample $k \in \{1,\dots,~K\}.$ 

In practice, the emphasis may lie in identifying treatment hierarchy questions for which an answer ``true" has high credibility, i.e. those $\mathcal Q$ for which $\ep(\mathcal Q) \geq \tau$, where $\tau$ is a pre-specified credibility threshold (e.g., 95\%). We refer to such cases as ``highly credible binary treatment hierarchy questions". In this section, we introduce efficient algorithms designed to systematically search through all possible binary treatment hierarchy questions of the classes defined in Section \ref{s:binQuest} and generate a catalog of all the highly credible ones. Guidance on choosing values for $\tau$ and $K$ is provided in Section \ref{sm:3_tau_K_guidance} of the Supplementary Materials.

\subsection{Uncovering Ranked and Unranked Permutations and Combinations}\label{s:algo1} \par

Due to the similar nature of the arrangement questions from Sections \ref{s:rankPerm}, \ref{s:perm}, \ref{s:rankComb}, and \ref{s:comb}, a single algorithm can be applied to first determine all observed ranked permutations, and then the other treatment hierarchy question types are determined through transformations of this output. Without loss of generality, we demonstrate this algorithm by making use of a toy example involving $K = 1,000$ simulated samples from a multivariate normal distribution with mean vector $(0.25, 0.95, 1.25, 1.5)'$, variances (0.025, 0.150, 0.025, 0.025), and covariances 0.010, which represents the joint posterior distribution of the relative effects of treatments B, C, D, E  vs. A. Smaller values are preferred, such that relative effects vs. A that are less than 0 indicate that the treatment is better than A. A matrix of hierarchies observed in each sample is the required input, where the column headers represent the ranks of each treatment and the row headers represent the index of each sample ($M_1$ in Section \ref{sm:4_1algo1} of the Supplementary Materials). Figure \ref{fig:algo1} illustrates the objects and outputs created by the single algorithm applied to this matrix of hierarchies.

\begin{figure}
\begin{center}
\includegraphics[height=0.9\textheight]{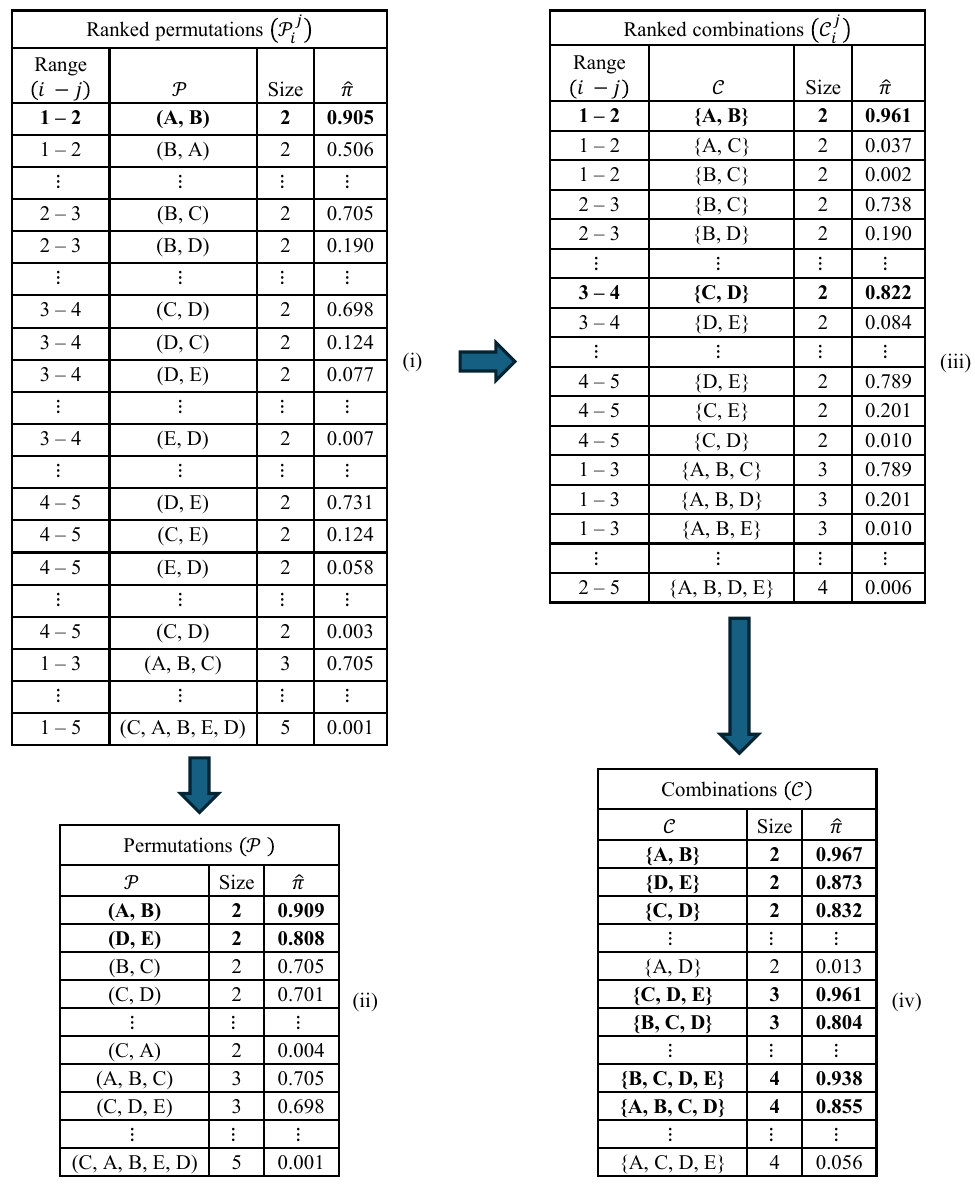}
\caption{Sample of objects produced by algorithm used to uncover ranked and unranked permutations and combinations in a toy example consisting of five treatments A, B, C, D, E. Treatment hierarchy questions that are credible at the $\tau = 0.80$ threshold are denoted in \textbf{bold}. Notation: $\mathcal{C}$ = combination, $\mathcal{C}_i^j$ = combination ranked $i$ through $j$, $\mathcal{P}$ = permutation, $\mathcal{P}_i^j$ = permutation ranked $i$ through $j$, $\ep$ = empirical probability.}
\label{fig:algo1}
\end{center}
\end{figure}

First, a list of all $N^{tot}$ unique observed ranked permutations from size 2 to $n$ is compiled with worst case time complexity $O(n^3K^2)$. Their empirical probabilities ($\ep$) are calculated, requiring $N^{tot}K$ evaluations (Figure \ref{fig:algo1}i). Here, $N^{tot}$ satisfies $\binom{n}{2}\leq N^{tot} \leq \binom{n}{2}K$, where the lower bound is achieved when all rows in $M_1$ are identical, and the upper bound occurs when all are distinct without overlap in their ranked permutations. Hence, noting that $K$ can remain modest for large values of $\tau$'s (Section \ref{sm:3_tau_K_guidance} of the Supplementary Materials), the time complexity of our approach represents a substantial gain compared to $O(n!K)$ for the brute force approach. Finally, all ranked permutations that meet the credibility threshold are reported. Only one ranked permutation meets a credibility threshold of $\tau = 0.80$ used in this demonstration: $(\text{A, B})_1^2$. 

To obtain the three other treatment hierarchy question types, simple aggregation operations which are not computationally expensive are applied to the ranked permutations and hence the impact on time complexity is kept minimal. We move onto unranked permutations by grouping object (i) in Figure \ref{fig:algo1} by permutation, summing the empirical probabilities of observed permutations regardless of their ranks, resulting in object (ii). Two credible unranked permutations are reported: (A, B) and (D, E). Note that the unranked permutation (D, E) is credible at the $\tau = 0.80$ threshold as a result of summing its empirical probabilities across all observed rank ranges in object (i): $\ep$(D, E) = $\ep$(D, E)$_3^4$ + $\ep$(D, E)$_4^5$ = 0.077 + 0.731 = 0.808. 

Next, we return to object (i) in Figure \ref{fig:algo1} to tabulate the observed ranked combinations and their empirical probabilities. This is done by sorting the permutation strings in alphabetical order to turn them into combination strings (that do not necessarily reflect the internal rank order of the treatments), and then grouping the output by combination strings and their ranks. This is only done for strings up to size $n-1$ as (ranked or unranked) combinations of size $n$ are not informative. This results in object (iii). Credible ranked combinations are then reported: $\{\text{A, B}\}_1^2, \{\text{C, D}\}_3^4, \{\text{C, D, E}\}_3^5, \{\text{B, C, D}\}_2^4, \{\text{B, C, D, E}\}_2^5, \{\text{A, B, C, D}\}_1^4$. Note there are six credible ranked combinations compared to one credible ranked permutation. As the order between the treatments within combinations does not matter, the empirical probabilities of the grouped treatments are expected to be larger for combinations than permutations. For example, $\ep$(C, D)$_3^4$ = 0.698, while $\ep$\{C, D\}$_3^4$ = $\ep$(C, D)$_3^4$ + $\ep$(D, C)$_3^4$  = 0.698 + 0.124 = 0.822.

Finally, similar to unranked permutations, to tabulate the observed unranked combinations, we group object (iii) in Figure \ref{fig:algo1} by combination, summing the empirical probabilities of observed combination regardless of their ranks. This results in object (iv). The credible unranked combinations include one additional group of treatments compared to the credible ranked combinations: $\{$D, E$\}$. Again, this is a result of summing the empirical probabilities over the rank ranges for which this combination was observed in object (iii) of Figure \ref{fig:algo1}: $\ep$\{D, E\} = $\ep$\{D, E\}$_3^4$ + $\ep$\{D, E\}$_4^5$  = 0.084 + 0.789 = 0.873.

\subsection{Uncovering Partial Hierarchies}\label{s:algo2} \par

To uncover the partial hierarchy features from Section \ref{s:poset}, a second algorithmic approach must be developed as the treatments do not have to be consecutive. The first difference lies in the way we utilize the samples from the joint posterior distribution of the relative effects. Here, we directly use the sampled relative effects, which are inputted as a matrix with treatments A through E as the column headers and iteration numbers as the row headers ($M_2$ in Section \ref{sm:4_2algo2} of the Supplementary Materials). 

First, all $n(n-1)$ possible partial hierarchies of size 2 are determined based on the treatment names given in $M_2$. Next, the empirical probability of each size-2 partial hierarchy is determined based on whether the difference in each treatment's relative effect is greater than or equal to the MID, which has been set to 0 in this demonstration: 
\begin{align}
    \begin{array}{cccccccccc} 
        h(\mathcal P, 0) & \ep & h(\mathcal P, 0) & \ep & h(\mathcal P, 0) & \ep & h(\mathcal P, 0) & \ep & h(\mathcal P, 0) & \ep \\
        \textbf{A} > \textbf{B} & \mathbf{0.942} & \text{B} > \text{A} & 0.058 & \text{C} > \text{A} & 0.006 & \text{D} > \text{A} & 0.000 & \text{E} > \text{A} & 0.000 \\
        \textbf{A} > \textbf{C} & \mathbf{0.994} & \textbf{B} > \textbf{C} & \mathbf{0.961} & \text{C} > \text{B} & 0.039 & \text{D} > \text{B} & 0.000 & \text{E} > \text{B} & 0.000 \\
        \textbf{A} > \textbf{D} & \mathbf{1.000} & \textbf{B} > \textbf{D} & \mathbf{1.000} & \text{C} > \text{D} & 0.792 & \text{D} > \text{C} & 0.208 & \text{E} > \text{C} & 0.087 \\
        \textbf{A} > \textbf{E} & \mathbf{1.000} & \textbf{B} > \textbf{E} & \mathbf{1.000} & \textbf{C} > \textbf{E} & \mathbf{0.913} & \textbf{D} > \textbf{E} & \mathbf{0.932} & \text{E} > \text{D} & 0.068
    \end{array}
    \label{eq:PH2}
\end{align}
Then, all size-2 partial hierarchies that meet the credibility threshold are recorded for output. At $\tau = 0.80$, the credible size-2 partial hierarchies are denoted in \textbf{bold} above. 

The credible size-2 partial hierarchies in object (\ref{eq:PH2}) above serve as the basis for building larger partial hierarchies that are potentially credible. To do this, we use two fundamental results on the monotonicity of probability. Let $T_{\{i\}} \in \mathcal{T}$ denote the $i^{th}$ ordered treatment in a partial hierarchy, $i = {1, 2, ..., k\leq n-1}$. First, $\text{if }\ep(T_{\{1\}} > T_{\{2\}} > ... > T_{\{k\}} > \text{MID}) \geq \tau, \text{then }\ep(T_{\{1\}} > T_{\{2\}} > ... > T_{\{k-1\}} > \text{MID}) \geq \tau.$ Based on this result, we know that the credible partial hierarchies of size $k$ must be an extension of the partial hierarchies of size $k-1$ that are credible. Also, $\text{if } \ep(T_{\{1\}} > T_{\{2\}} > ... > T_{\{k\}} > \text{MID}) \geq \tau, \text{then }\ep(T_{\{k-1\}} > T_{\{k\}} > \text{MID}) \geq \tau$. By restricting the expanded partial hierarchies (i.e., those with treatments added to the right) to those that satisfy the above results, the efficiency of the credibility assessment improves. For example, while $\frac{5!}{(5-3)!} = 60$ size-3 partial hierarchies may be formed based on {A, B, C, D, E}, we are able to restrict our attention to 15 size-3 partial hierarchies in our toy example. This is because we know expansions of size-2 partial hierarchies that are not credible in object (\ref{eq:PH2}) will also not be credible, reducing the number of size-3 partial hierarchies to consider by 33. We further ignore 20 size-3 partial hierarchies that involve credible size-2 partial hierarchies expanded by non-credible size-2 partial hierarchies. For example, extensions of size-2 partial hierarchies ending in E (e.g. A $>$ E) would lack credibility when expanded to size-3 (e.g. A $>$ E $>$ D) because there are no credible size-2 partial hierarchies beginning with E in object (\ref{eq:PH2}). So, the list of potentially credible size-3 partial hierarchies begins with the first size-2 partial hierarchy that may be expanded: A $>$ B. We append the credible size-2 partial hierarchies beginning with B in object (\ref{eq:PH2}) (i.e., B $>$ C, B $>$ D, B $>$ E) to the right. Then, we continue this process for every other eligible credible size-2 partial hierarchy in object (\ref{eq:PH2}) and end up with the following list of potentially credible size-3 partial hierarchies: A $>$ B $>$ C; A $>$ B $>$ D, A $>$ B $>$ E; A $>$ C $>$ E; A $>$ D $>$ E; B $>$ C $>$ E; B $>$ D $>$ E.

The algorithm then computes the empirical probabilities of each of these potentially credible size-3 partial hierarchies and reports the ones that meet the credibility threshold. In this case, all of them were credible at a threshold of 0.80:
\begin{align}
    \begin{array}{cccccccccc} 
        h(\mathcal P, 0) & \ep & h(\mathcal P, 0) & \ep & h(\mathcal P, 0) & \ep \\
        \textbf{A} > \textbf{B} > \textbf{C} & \mathbf{0.905} &  \textbf{A} > \textbf{C} > \textbf{E} & \mathbf{0.907} &  \textbf{B} > \textbf{C} > \textbf{E} & \mathbf{0.874}  \\
        \textbf{A} > \textbf{B} > \textbf{D} & \mathbf{0.942} & \textbf{A} > \textbf{D} > \textbf{E} & \mathbf{0.932} & \textbf{B} > \textbf{D} > \textbf{E} & \mathbf{0.932}  \\
        \textbf{A} > \textbf{B} > \textbf{E} & \mathbf{0.942} & & & &  
    \end{array}
    \label{eq:PH3}
\end{align}

We then repeat this process to determine credible partial hierarchies of size 4, expanding the credible size-3 partial hierarchies in object (\ref{eq:PH3}) by appending to the right credible size-2 partial hierarchies in object (\ref{eq:PH2}). This leads to two size-4 partial hierarchies that could be potentially credible: A $>$ B $>$ C $>$ E and A $>$ B $>$ D $>$ E. These both end up being credible at $\tau = 0.80$ ($\boldsymbol{\ep}\textbf{(A} > \textbf{B} > \textbf{C} > \textbf{E) = 0.823}, \boldsymbol{\ep}\textbf{(A} > \textbf{B} > \textbf{D} > \textbf{E) = 0.879})$. The algorithm then concludes as credible partial hierarchies of the maximum size $n=5$ are credible ranked permutations, which would be detected in the algorithm described in Section \ref{s:algo1}. 

The number of empirical probabilities that need to be computed using this approach decreases as the threshold $\tau$ increases and the degree of overlap between the marginal posterior distributions of the relative treatment effects decreases. Let $N_s^\tau(\text{MID})$ be the number of credible size-$s$ partial hierarchies at threshold $\tau$ and a specific MID, then the total number of required calculations at size ($s+1$) is bounded above by ${N_s^\tau(\text{MID})\times (n-s) \times K}$. In most NMAs, this will provide a significant gain in computation time over the brute force approach, which has a time complexity of $O(n2^nK)$, since higher thresholds are typically of greatest interest and many NMAs lack sufficient evidence (i.e., exhibit substantial overlap in marginal posterior distributions) to yield a large $N_s^\tau(\text{MID})$. In addition, our algorithm only proceeds with expanding the size of partial hierarchies from $s$ to $s+1$ if $N_s^\tau(\text{MID}) \neq 0$ ; this means it may stop at a size less than $n-1$, hence providing an additional gain in efficiency compared to the brute force approach.

\subsection{Uncovering Ranks for Individual Treatments} \label{s:algo3} \par

Rather than evaluating all $T_\mathcal R$ treatment hierarchy questions individually at a credible threshold $\tau$, we propose that all credible questions at a pre-specified $\tau$ can be efficiently addressed by computing the highest (posterior) density region (HDR) of ranks for each treatment (see e.g. \citet{gelman2014}). The HDR set is the subset of ranks with the smallest possible cumulative frequency that is at least equal to $\tau$, where each rank in the subset has a higher density than any rank not included in the subset. 

To motivate HDR as the preferred approach, consider a treatment in an NMA with posterior ranking probabilities of 0.001, 0.001, 0.005, 0.15, 0.4, 0.3, 0.14, 0.002, 0.001 for ranks 1 through 9, respectively. At the threshold $\tau = 0.95$, the HDR consists of ranks $\{4, 5, 6, 7\}$, that is the smallest set of ranks whose combined probability meets or exceeds the threshold. While other sets such as $\{1, 2, 3, 4, 5, 6, 7\}$, the top 7 ranks in a plot of the cumulative rankings (surface under the cumulative ranking [SUCRA] plot) \citep{salanti2011}, also exceed the 95\% threshold, they can be misleading by overstating top rankings that are, in fact, highly improbable.

As a second example, consider a treatment with a multi-modal posterior ranking distribution, which can occur when the number of studies across comparisons is unbalanced \citep{kibret2014}. Suppose the posterior ranking probabilities are 0.55, 0.03, 0.01, 0.41 for ranks 1 through 4, respectively. At the threshold $\tau = 0.95$, the HDR consisting of ranks 1 and 4 is the only meaningful choice, as it captures the two dominant modes of the distribution.

The HDR method is computationally efficient, scaling well even in large treatment networks, while accurately reflecting the concentration of ranking probabilities, as illustrated above. Furthermore, when a treatment’s ranking distribution is skewed toward the top or bottom ranks, the HDR will capture this pattern, emphasizing the top or bottom ranks and aligning with insights sought from a SUCRA plot. Moreover, if a single rank holds a posterior probability of at least $\tau$, the HDR will consist solely of that rank, as is desirable.

Since the computation of HDR is well-established, we have demonstrated its application to the toy example in Section \ref{sm:5_algo3} of the Supplementary Materials.

\subsection{Trimming Redundant Hierarchy Questions}\label{s:trim}

The methods described in Sections \ref{s:algo1}-\ref{s:algo3} may uncover multiple hierarchies with overlapping information. For example, consider the algorithmic approach described in Section \ref{s:algo1} which first uncovers all credible ranked permutations of treatments at a threshold $\tau$. Credible unranked permutations, ranked combinations and unranked combinations involving the same treatments would also be uncovered at the same threshold by collapsing information provided by the ranked permutations, but the information provided by the former is less precise and may therefore be perceived as redundant. To present the most concise information among highly credible binary treatment hierarchy questions to decision makers, we recommend trimming the redundancies outlined in Section \ref{sm:2_trim} of the Supplementary Materials.

\section{Empirical Illustrations} \label{s:empIll}

\begin{figure}[ht]
\begin{center}
\includegraphics{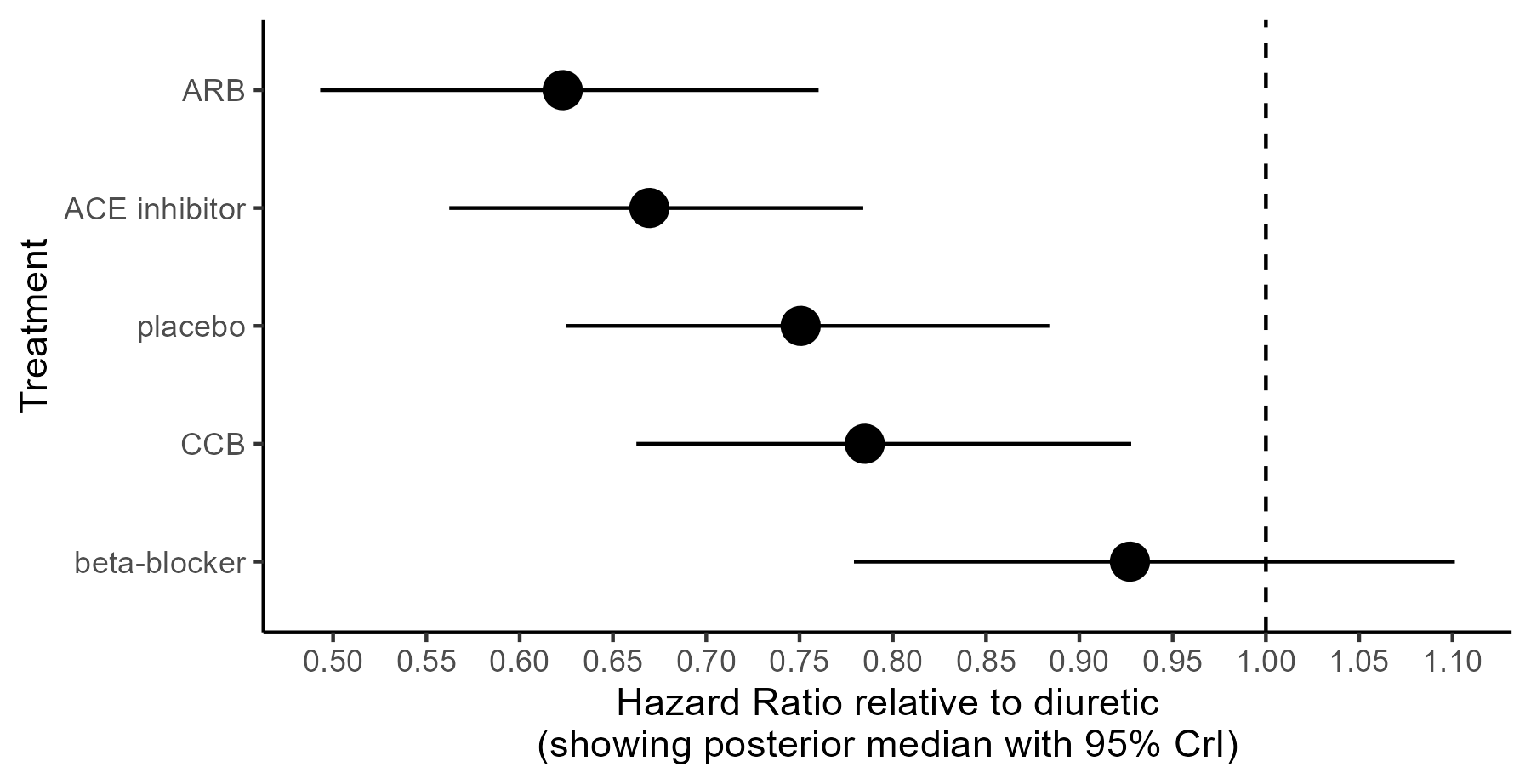}
\caption{Forest plot of the estimated effects of antihypertensive classes and placebo vs. diuretics in terms of incident diabetes based on an NMA \citep{elliott2007, dias2018}. Abbreviations: ACE = angiotensin-converting enzyme, ARB = angiotensin receptor blockers, CCB = calcium channel blockers, CrI = credible interval.}
\label{fig:diabetes_fp}
\end{center}
\end{figure}

To illustrate the features of the proposed algorithms, we applied them to an NMA dataset comparing the effectiveness of five antihypertensive drug classes (angiotensin-converting enzyme [ACE] inhibitors, angiotensin receptor blockers [ARB], beta-blockers, calcium channel blockers [CCB], and diuretics) and placebo in terms of the incident diabetes \citep{elliott2007}. The NMA model fitted in \texttt{BUGSnet} \citep{bugsnet} is described in Section \ref{sm:6_diabetes} of the Supplementary Materials. A forest plot of the estimated relative effects (hazard ratios) of the various classes vs. diuretics is presented in Figure \ref{fig:diabetes_fp}. Based solely on their point estimates, a hierarchy of all $n=6$ treatments may be $($ARB, ACE inhibitor, placebo, CCB, beta-blocker, diuretic$)_1^6$. However, there is some overlap in marginal posterior distributions of the relative effects between all treatments, reflected by the credible intervals (CrIs), which adds some uncertainty to this hierarchy. We note that the highest degree of overlap in 95\% CrIs is between placebo and CCB. In addition, the 95\% CrI for beta-blockers overlaps with the null effect, suggesting there is not enough evidence to suggest a difference in the effects of beta-blockers and diuretics.

Preparing the sampled relative effects for the proposed algorithms took approximately 0.09 seconds. Generating the credible (redundant and non-redundant) treatment hierarchy questions in Sections \ref{s:algo1}, \ref{s:algo2}, and \ref{s:algo3} took approximately 0.20, 0.20, and 0.01 seconds, respectively, on a laptop with an Intel Core 7 processor and 16.2 GB of RAM. The algorithms revealed no credible ranked permutations nor unranked permutations, but two ranked combinations, two unranked combinations, 17 partial hierarchies ranging from size two to three, and six HDRs encompassing two to three ranks were credible at a  threshold of $\tau = 0.95$ (Supplementary Materials, Table \ref{tab:diab_red_table}). None of the unranked combinations added information beyond what was available and credible in the ranked combinations, and thus the former were redundant. All size-two partial hierarchies were made redundant by the size-three partial hierarchies. In addition, the HDRs beta-blocker$_{\{5-6\}}$ and diuretics$_{\{5-6\}}$ were made redundant by the ranked combination \{beta-blocker, diuretic\}$_5^6$. The final number of non-redundant hierarchies to consider is 12: two ranked combinations, six partial hierarchies, and four HDRs (Table \ref{tab:diabetes_table}). 

\begin{table}[ht]
  \centering
    \begin{tabular}{lll}
    \hline
        Type & Treatment Hierarchy Question & $\ep$ \\
        \hline
        Ranked Combination & $\{$ACE inhibitor, ARB, CCB, placebo$\}_1^4$ & 0.9773 \\
        & $\{$beta-blocker, diuretic$\}_5^6$ & 0.9773 \\
        \hline
        Partial Hierarchy & ARB $>$ CCB $>$ diuretic & 0.9920 \\
        at MID = 0& ARB $>$ CCB $>$ beta-blocker & 0.9857 \\
        & ARB $>$ placebo $>$ diuretic & 0.9793 \\
        & ACE inhibitor $>$ CCB $>$ diuretic & 0.9773 \\
        & ACE inhibitor $>$ CCB $>$ beta-blocker & 0.9710 \\
        & ARB $>$ placebo $>$ beta-blocker & 0.9707 \\
        \hline
        HDR & ARB$_{\{1-2\}}$ & 0.9790 \\
        & ACE inhibitor$_{\{1-3\}}$ & 0.9940 \\
        & placebo$_{\{2-4\}}$ & 0.9873 \\
        & CCB$_{\{3-4\}}$ & 0.9703 \\
        \hline
    \end{tabular} 
    \caption{Non-redundant hierarchy questions regarding the effectiveness of antihypertensive drugs and placebo in terms of incident diabetes, which exceed credibility threshold 0.95. Abbreviations: ACE = angiotensin-converting enzyme, ARB = angiotensin receptor blockers, CCB = calcium channel blockers, HDR = high density region, MID~=~minimally important difference. Notation: $\ep$ = empirical probability.}
    \label{tab:diabetes_table}
\end{table}

\begin{figure}[ht]
\begin{center}
\includegraphics[width=1.0\textwidth]{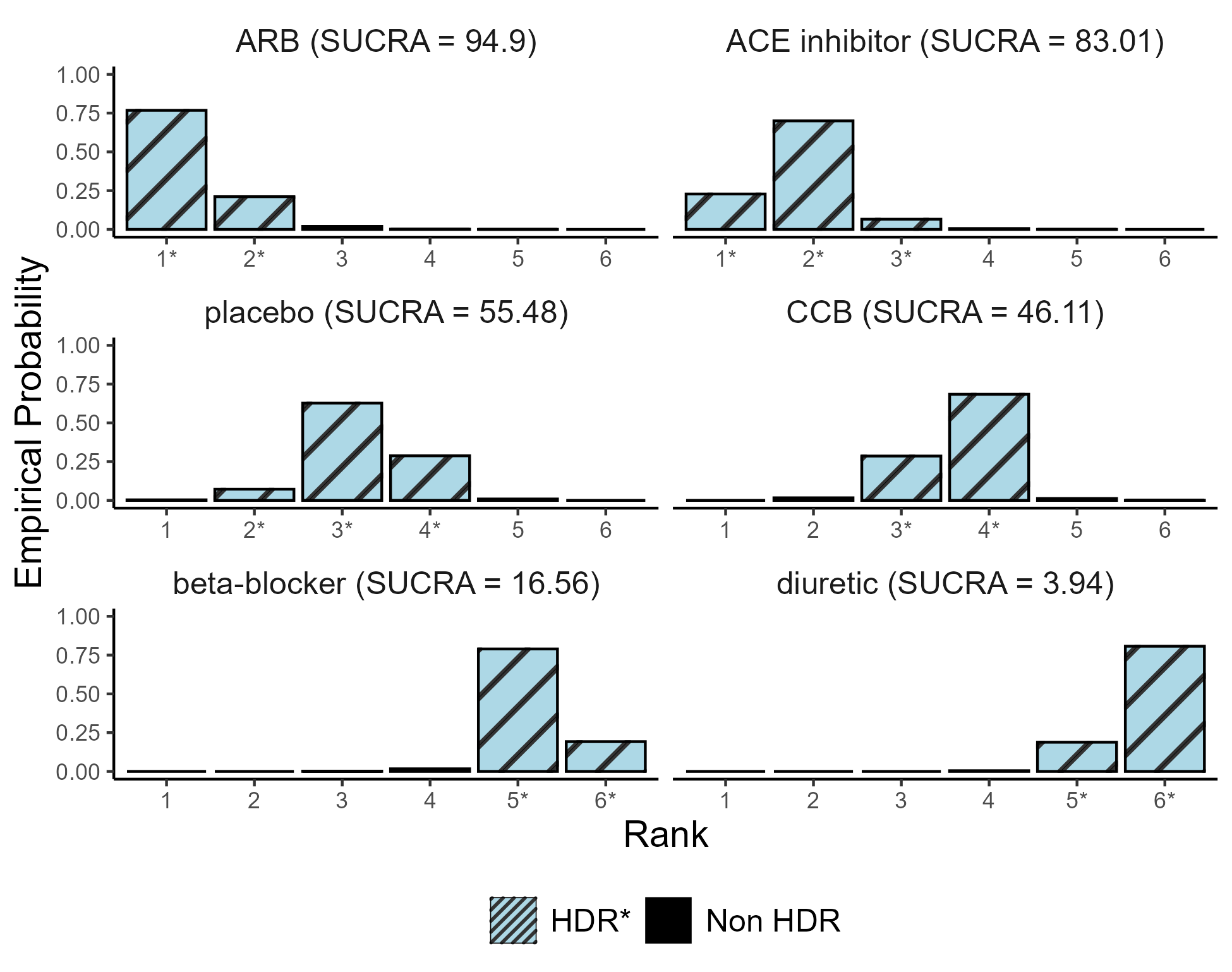}
\caption{Rankograms with HDRs for each treatment in diabetes network based on an NMA \citep{elliott2007, dias2018}. Blue bars with black stripes and asterisks indicate the ranks belonging to the HDR set. Abbreviations: ACE = angiotensin-converting enzyme, ARB = angiotensin receptor blockers, CCB = calcium channel blockers, HDR = high density region, SUCRA = surface under the cumulative ranking curve.}
\label{fig:diabetes_hdr}
\end{center}
\end{figure}

In line with the uncertainty of the speculated hierarchy based on the forest plot (Figure \ref{fig:diabetes_fp}), the evidence does not support a complete hierarchy question involving all six treatments (Table \ref{tab:diabetes_table}). However, smaller treatment hierarchy questions are credible. It is very likely that ACE inhibitors, ARBs, placebo, and CCBs occupy the top four ranks, while beta-blockers and diuretics rank either 5th or 6th. The groupings of these treatments correspond with the observation that the 95\% CrIs of the top four ranking treatments do not overlap with the null effect and that there is more overlap in their 95\% CrIs between each ordered pair of treatments in this group (ARB-ACE inhibitor, ACE inhibitor-placebo, placebo-CCB) compared to that between CCBs and beta-blockers (Figure \ref{fig:diabetes_fp}). The credible partial hierarchies add some insight in terms of the ordering of some treatments within the two ranked combinations: ARB and ACE inhibitors are likely to be better than CCB and placebo, but the ranking between ARB and ACE inhibitors, CCB and placebo, as well as beta-blockers and diuretics is less clear. For completeness, the rankograms with the HDRs overlayed are presented in Figure \ref{fig:diabetes_hdr}, which further highlights the uncertainty of each individual treatment's rank. While all treatments had a considerably large empirical probability of ranking a specific rank, none of these probabilities met the credibility threshold of $\tau = 0.95$, and hence their HDRs encompass multiple ranks. Overall, in this example, the partial hierarchies were most informative on clear orderings within the true underlying treatment hierarchy.

With the same number of samples, a sensitivity analysis using $\tau^* = 0.95 - 2\sqrt{0.95(0.05)/500} = 0.9305$ yielded two additional partial hierarchies (ACE inhibitor $>$ placebo $>$ diuretic; ACE inhibitor $>$ placebo), with minimal change in computation time. These additional partial hierarchies were likely revealed due to the degree of overlap between the marginal posterior distributions of the relative effects for ACE inhibitor and placebo vs. diuretics. 

Another empirical illustration of the proposed algorithms is presented in Section \ref{sm:7_cbtdep} of the Supplementary Materials using an NMA that compares cognitive behavioral therapies (CBT) and controls for depression \citep{cuijpers2019}. This example has a similar number of treatments as the diabetes example, but with a different pattern of overlap in the marginal posterior distributions of the relative treatment effects, resulting in some intriguing credible treatment hierarchy questions including a ranked permutation.

\section{Discussion}\label{s:discuss}

Questions often motivate some form of research. In the context of medical treatments, clinical research questions are typically framed around the overarching question, ``What is the optimal treatment?", where the criteria to determine optimality is explicitly defined. Evidence synthesis researchers have recently laid the foundation for identifying and tying treatment hierarchy metrics to relevant, detailed decision questions \citep{salanti2022_hierachyq, papakonstantinou2022}. In this paper, we build on this work by uncovering the vast number of treatment hierarchy questions that may be answered through NMA. More specifically, we describe six types of binary treatment hierarchy questions which may be answered through empirical probabilities of ranked permutations, permutations, ranked combinations, combinations, partial hierarchies, and HDRs. We developed three efficient algorithms to identify the credibility of these treatment hierarchy questions. Noting the potential overlap in the information provided by these descriptive summaries, we also identified 18 types of redundant treatment hierarchy questions that may be trimmed to provide a concise summary of the evidence to decision makers. To help facilitate the application of the hierarchy algorithms and redundancy checks described in this paper, we have developed an R package, \texttt{hphq}, which is available on GitHub (https://github.com/caitlin-h-daly/hphq).

We used two networks of evidence to illustrate how the proposed algorithms may be used to uncover credible and non-redundant hierarchy questions. In the diabetes example, there was some overlap in the marginal posterior distributions (and hence credible intervals) of the relative treatment effects vs. a common treatment, while in the depression example, there were clear instances where there was almost no overlap. A complete and fully informative hierarchy involving all treatments (i.e., $\mathcal{P}_1^n$) was not credible at the $\tau = 0.95$ threshold in either example due to the considerable overlap between the posterior distributions of some treatments' relative effects. Nevertheless, the minimal overlap between the posterior distributions of other treatments' relative effects led to the uncovering of smaller credible treatment hierarchy questions which collectively provided some insight into each treatment's position in the hierarchy. As the degree of overlap between posterior distributions (and hence credible intervals) decreased, the credible treatment hierarchy questions were more informative. When there is lots of overlap between the credible intervals (or confidence intervals in the case of a frequentist NMA) of the relative treatment effects, as is the case in an example involving 18 anti-depressants showcased in \citet{papakonstantinou2022}, our approach will likely demonstrate that no in-depth treatment hierarchies questions are credible at a high threshold, highlighting that any conclusive statements on treatment hierarchies should be avoided or interpreted with caution. Applying our algorithms to this anti-depressant example using 500 independent samples from an estimated joint distribution of relative effects in a frequentist NMA, the only credible hierarchy questions identified at the $\tau = 0.95$ threshold were 35 partial hierarchies of size 2 (i.e. pairwise comparisons) and HDRs for each treatment ranging from size 7 to 18. In addition, we note that, compared to the diabetes and CBT depression examples, the computation time for the first algorithm (Section \ref{s:algo1}) was longer (3.81 seconds); this can be explained by the larger number of treatments in the network, which leads to an increased number of observed ranked permutations to be tabulated by this algorithm. As was the case for the diabetes and CBT depression examples, the computation times for the second and third algorithms (Sections \ref{s:algo2} and \ref{s:algo3}) were less than 1 second.

The questions we addressed uncover evidence about the fundamental truths concerning the treatment hierarchy, whether that be in part or in whole, under a specific decision problem. We did not examine questions such as ``Does treatment A have the highest SUCRA value/mean rank?" or ``Is the median rank value for A less than 3?" because such questions pertain to the distribution of statistics rather than directly addressing the true underlying treatment hierarchy. These questions focus on the distribution of statistical metrics and their answers are inherently data-dependent, whereas the questions we considered aim to uncover a unique, true underlying answer about the treatment hierarchy. This is not to suggest that questions such as the two defined above lack relevance. However, one could argue that a decision maker asking either of the two questions is ultimately seeking to determine whether A is among the top-performing treatments. While historically these questions have been addressed through metrics such as pairwise differences, mean ranks, median ranks, or SUCRA values \citep{petropoulou2017}, we encourage the approach proposed by \citet{papakonstantinou2022}, where empirical probabilities are used to draw inferences on the true underlying treatment hierarchy. 

Our approach requires the computation of empirical probabilities for multiple hierarchy questions. This may draw criticism if viewed as multiple hypothesis testing. An adjustment factor applied to the credibility threshold akin to the Bonferonni correction would be too conservative to produce any meaningful output due to the large number of treatment hierarchy questions. More thought to account for Type I error may be needed. The current proposed approach minimizes the number of treatment hierarchy questions for consideration by internally trimming redundant questions. Note that our list of potential redundancies is not exhaustive. If further redundancies become apparent to us in future research, we will update our \texttt{hphq} R package to incorporate these options.

Additionally, we have not developed an algorithm which uncovers credible compound treatment hierarchy questions consisting of intersections of our hierarchy questions. For example, ``Are A and B among the top 2 treatments and are C, D, E ranked consecutively anywhere else in the hierarchy?". Such questions are likely motivated by a priori knowledge and could be investigated individually using the \texttt{nmarank} R package developed alongside the methods proposed by \citet{papakonstantinou2022}. Note it is only worth investigating compound questions if each question forming the compound is itself credible. Otherwise, the intersection cannot be deemed credible. Therefore, our approach is beneficial for narrowing down compound questions to only the potentially credible ones. Note that, compounding credible, consecutive, hierarchy questions of the same type (e.g., \{A, B\}$^1_2 ~\cap$ \{C, D, E\}$_3^5$) will not result in credible questions as they would have been outputted in the first place.

In the early use of NMA methodology, the reporting of ranking statistics were recommended by professional bodies \citep{jansen2011}. However, overtime, reliance on ranking statistics to interpret NMA has been discouraged for various reasons. A valid and important criticism is that differences in treatment ranks do not reflect meaningful differences between treatments' performances \citep{mills2012, mbuagbaw2017, trinquart2016}. Our algorithms currently only allow for the consideration of MIDs when uncovering partial hierarchies. In the future, we plan to incorporate MIDs in other types of binary treatment hierarchy questions. An additional criticism arises when treatments with very uncertain or multimodal ranking distributions have the highest probability of ranking best \citep{kibret2014, jansen2011}. Our proposed methods account for this through the reporting of empirical probabilities of HDRs, instead of focusing on single treatment ranks. Large credibility thresholds (e.g., $\tau \geq 0.95$) would also likely exclude such cases. Finally, hierarchy summaries have also been criticized for overlooking concerns regarding the quality of the evidence \citep{mbuagbaw2017}. We strongly encourage the outputs to be interpreted in light of such quality concerns, and should be used as a supplementary summary of the NMA evidence, rather than the primary summary. Nevertheless, our proposed approach may be viewed as an alternative to ranking statistics, and offers the advantage of only reporting clear hierarchy features that are strongly supported by the evidence. When such clarity exists at large credibility thresholds, it seems intuitive that this occurs for parts of the hierarchy that are less subject to bias. As such, at large credibility thresholds, highly credible treatment hierarchy questions may be more trustworthy than ranking statistics.

As noted in the illustrative examples, at high thresholds, the methods developed in this paper will likely only uncover credible treatment hierarchy questions when there is minimal overlap between the marginal posterior distributions of the relative treatment effects. Many NMAs lack sufficient evidence for our approach to yield insights beyond those provided by league tables and rankograms. While searching for suitable examples to illustrate our approach, we discovered several NMAs with high precision of treatment hierarchies (POTH \citep{wigle2025}) did not provide highly credible treatment hierarchy questions apart from small partial hierarchies and HDRs. This is not a shortcoming of our approach, but a valuable reminder that confidence should not be placed in treatment hierarchies produced from these NMAs. As certainty in NMAs increases with the accumulation of evidence over time, more treatment hierarchy questions will be identified as highly credible.

We conclude with a reminder of the importance of interpreting treatment hierarchy questions within the decision problem addressed by the evidence. That is, the conditions (e.g., population, outcome, setting) they are applicable to. In addition, we note that while the methods proposed here have been developed in the context of comparing medical treatments based on evidence synthesized through NMA, they are applicable to any decision problem involving three or more options and for which there is joint evidence on their relative performance. The key data used to inform all our algorithms are simply derived from samples of the joint distribution of the relative treatment effects. 

\section*{Acknowledgements}

This research was supported by an NSERC Discovery Grant (RGPIN-2019-04404).\vspace*{-8pt}

\bibliographystyle{plainnat}
\bibliography{arXiv/arXiv_references}  

\newpage
\setcounter{page}{1}
\setcounter{section}{0}
\setcounter{table}{0}
\setcounter{figure}{0}
\setcounter{equation}{0}

\makeatletter
\renewcommand \thesection{S\@arabic\c@section}
\renewcommand \thetable{S\@arabic\c@table}
\renewcommand \thefigure{S\@arabic\c@figure}
\renewcommand \theequation{S\@arabic\c@equation}
\makeatother

\begin{center}
\Large{\textbf{Supplementary Material for ``Uncovering All Highly Credible Binary Treatment Hierarchy Questions in Network Meta-Analysis''}} \\ [2ex]
by \\ [2ex]
\large{Caitlin H Daly, Chloe Tan, Audrey B\'{e}liveau}
\end{center} \bigskip

\section{Alignment with Established NMA Reporting Tools}\label{sm:1_reporttools}

Numerous established reporting tools exist within NMA, and it is essential to clarify how the treatment hierarchy questions outlined in Section \ref{s:binQuest} of the main manuscript align with them.

League tables and forest plots \citep{florez2024}, which assess pairwise differences between treatments, are closely aligned with partial hierarchies of size 2 (Section \ref{s:poset} of the main manuscript). Typically, evidence of a difference between two treatments A and B is determined by assessing whether the $100p\%$ credible interval (CrI) or confidence interval (CI) of their relative effect includes the null effect. When the interval indicates a difference between A and B at the $100p\%$ credibility level, the associated partial hierarchy— $h$((A,~B),~0) if the observed effect favors A over B or $h$((B,~A),~0) if it favors B over A— will have empirical probability of at least $p+(1-p)/2$. The second term arises because league tables and forest plots evaluate two-sided comparisons, whereas size-2 partial hierarchies are one-sided comparisons.

Rankograms \citep{salanti2011} report the empirical probability of each treatment taking each rank. These are reported within treatment hierarchy questions for individual treatments $T_\mathcal{R}$, where $T$ is any individual treatment in the NMA and $\mathcal{R}$ any individual rank (Section \ref{s:rankSets} of the main manuscript).

Plots of the cumulative rankings, often termed Surface Under the Cumulative Ranking curve [SUCRA] plots \citep{salanti2011}, illustrate, for each treatment, the probability its performance is among the top $m$ treatments, for $m=1, ..., n-1$. That is, the cumulative ranking probabilities. These are reported within treatment hierarchy questions for individual treatments $T_\mathcal{R}$, where $T$ is any individual treatment in the NMA and $\mathcal{R}$ is the set $\{1,\dots,~m\}$ (Section \ref{s:rankSets} of the main manuscript). Similarly, the probability a treatment's performance is among the bottom $m$ treatments, for $m=1, ..., n-1$ is evaluated via $\mathcal{R}=\{n-m+1,\dots,~n\}$.

As our focus is exclusively on binary treatment hierarchy questions pertaining to the true underlying treatment hierarchy, the questions we ask are not directly comparable to other popular NMA metrics such as median rank or mean rank (or equivalently SUCRA values or P-scores \citep{salanti2011, rucker2015}), which are statistical summaries specific to the observed data and not hierarchy questions in themselves. Therefore, we recommend reporting these metrics separately from the investigation of treatment hierarchy questions.

In summary, the classes of treatment hierarchy questions defined in Sections \ref{s:rankPerm} to \ref{s:rankSets} of the main manuscript not only capture major NMA reporting tools (league tables and rankograms) but also extend to a wider array of complex questions that are often neglected in standard NMA practice due to the lack of appropriate methodology. We address this methodological gap in Section \ref{s:algo_descriptions} of the main manuscript.

\clearpage

\section{Methods for Trimming Redundant Hierarchy Questions}\label{sm:2_trim}

To present the most concise information among highly credible binary treatment hierarchy questions to decision makers, we recommend trimming the redundancies outlined in this section and exemplified in Table \ref{t:redundant}.

Note if all types of redundancies are identified before trimming, then the order of identification does not matter. Otherwise, trimming of more informative hierarchy question types should only occur after the redundancy status of less informative hierarchy question types has been determined (i.e., combinations $<$ ranked combinations $<$ permutations $<$ ranked permutations; partial hierarchies with MID = 0 $<$ permutations). Regardless, to maximize efficiency, each type of redundancy detection should be restricted to treatment hierarchy questions that have not already been determined to be redundant. 

\begin{table}[hb]
    \centering
    \begin{tabular}{lccc}
    \hline
    & If this hierarchy question & & then this one is also\\
    & is highly credible & & highly credible and redundant\\
    \hline
    & Subset &  & Superset\\
    \hline
    1 & (A, B, C)$_1^3$  & $\implies$& (A, B)$_1^2$\\
    2 & (A, B, C) & $\implies$& (A, B)\\
    3 & $h$((A, B, C), MID) & $\implies$& $h$((A, B), MID) \\
    4 &  (A, B, C) & $\implies$& $h$((A, B, C), 0) \\
    5 & (A, B, C)$_1^3$ & $\implies$& (A, B, C)\\    
    6 & $\{$A, B, C$\}_1^3$ & $\implies$& $\{$A, B, C$\}$\\
    7 & (A, B, C) & $\implies$& $\{$A, B, C$\}$\\
    8 & (A, B, C)$_1^3$ & $\implies$& $\{$A, B, C$\}_1^3$\\
    9 & (A, B, C)$_1^3$ & $\implies$ & A$_{\{1\}}$, B$_{\{2\}}$, C$_{\{3\}}$ \\
    10 & $\{$A, B, C$\}_1^3$ & $\implies$ & A$_{\{1-3\}}$, B$_{\{1-3\}}$, C$_{\{1-3\}}$ \\
    \hline
    & Top/bottom partition block& & Bottom/top partition block\\
    \hline
    11 & $\{$A, B, C$\}_1^3$ & $\iff$ & $\{$D, E, F$\}_4^6$ \\
    12a & A$_{\{1\}}$ & $\iff$ & $\{$B, C, D, E, F$\}_2^6$ \\
    12b & F$_{\{6\}}$ & $\iff$ & $\{$A, B, C, D, E$\}_1^5$ \\
    13 & (A, B, C, D, E)$_1^5$ & $\implies$ & F$_{\{6\}}$ \\
    \hline
    & Tail partition blocks & & Middle partition block\\     
    \hline
    14 & A$_{\{1\}}~\cap $ F$_{\{6\}}$ & $\implies$ & $\{$B, C, D, E$\}_2^5$ \\
    15 & A$_{\{1\}}~\cap $ (E, F)$_5^6$ & $\implies$ & $\{$B, C, D$\}_2^4$ \\
    16 & (A, B)$_1^2 ~\cap $ (E, F)$_5^6$ & $\implies$ & $\{$C, D$\}_3^4$ \\
    \hline
    & Divided partition block & & Partition block \\
    \hline
    17 & (A, B)$_1^2 ~\cap $ \{C, D\}$_3^4$ & $\implies$ & \{A, B, C, D\}$_1^4$ \\
    18 & A$_{\{1\}} ~\cap $ \{B, C\}$_2^3$ & $\implies$ & \{A, B, C\}$_1^3$ \\
    \hline
    \end{tabular}
    \caption{Examples of redundancy types in a network of 6 treatments: A, B, C, D, E, F. The hierarchy questions on the right would be recommended for trimming. Notation: \{.\} = combination, \{.\}$_i^j$ = combination ranked $i$ through $j$, (.) = permutation, (.)$_i^j$ = permutation ranked $i$ through $j$, $h$(., MID) = partial hierarchy, $._{\{.\}}$ = high density region, $\implies$ = implies (one direction), $\iff$ = equivalent, $\cap$ = intersection.}
    \label{t:redundant}
\end{table}

\subsection{Trimming Supersets}\label{sm:2_1trimsup} \par

Among the credible hierarchy questions uncovered using the methods in Section \ref{s:algo_descriptions} of the main manuscript, some may encompass others as supersets. For example, (A, B)$_1^2$ is a superset of (A, B, D)$_1^3$ since it can be expressed as (A, B)$_1^2 = \bigcup_{t \in \mathcal{T}\backslash\{\text{A, B}\}}$ (A, B, $t$)$_1^3$. Similarly, another possible superset of (A, B, D)$_1^3$ is A$_{\{1\}}$. Our simplifying notations for treatment hierarchy questions focus on the treatments subject to constraints, which, counterintuitively, make supersets more concise in notation than subsets. Since supersets provide less detailed information than their subset counterparts, it can be valuable to identify all supersets in the uncovered credible hierarchies in order to trim them from the results. 

Generally, we classify subset-superset relationships into three categories. The first category is composed of relationships that happen within a particular type of treatment hierarchy question. These can arise within ranked permutations (Section \ref{s:rankPerm} of the main manuscript), permutations (Section \ref{s:perm} of the main manuscript), and partial hierarchies (Section \ref{s:poset} of the main manuscript) (rows 1-3 in Table \ref{t:redundant}). The previous example (A, B)$_1^2 \supset ($A, B, D)$_1^3$ demonstrates such a relationship within ranked permutations. More generally, for ranked permutations, we have $\mathcal{P}_i^j \supset \tilde{\mathcal{P}}_{k}^{l}$ when $\mathcal{P} \subset \tilde{\mathcal{P}}$ and $k\leq i<j\leq l$ with at least one ``$\leq$" being a strict inequality (row 1 in Table \ref{t:redundant}). Similar implications hold for unranked permutations (row 2 in Table \ref{t:redundant}) and for partial hierarchies (row 3 in Table \ref{t:redundant}). To identify all supersets, we compare each credible treatment hierarchy question to all other credible treatment hierarchy questions of the same type that are larger in size and determine whether they are supersets. For example, if we were examining the partial hierarchy A $>$ B, we would look through all other partial hierarchies with size $\geq 3$. If we detect A and B in that order in any of the larger partial hierarchies, then A $>$ B is labeled a superset.

The second subset-superset relationship category occurs across different hierarchy question types (rows 4-8 in Table \ref{t:redundant}). To identify these supersets, we compare hierarchy question types with those from a more informative hierarchy question type. Continuing with partial hierarchies, we compare all partial hierarchies with MID = 0 with permutations of the same treatments (row 4 in Table \ref{t:redundant}). Doing this after trimming within partial hierarchies gives us the full list of non-redundant credible partial hierarchy questions. Permutations involve an identical approach, where we first compare credible permutations amongst themselves, followed by comparing the permutations with ranked permutations of the same treatments (row 5 in Table \ref{t:redundant}). Combinations (Section \ref{s:comb} of the main manuscript) are compared with ranked combinations (row 6 in Table \ref{t:redundant}) and (unranked) permutations (row 7 in Table \ref{t:redundant}) of the same treatments, while ranked combinations (Section \ref{s:rankComb} of the main manuscript) are compared with ranked permutations (row 8 in Table \ref{t:redundant}). 

A third subset-superset relationship category is one that compares the union (superset) and intersection (subset) of multiple hierarchies answering the same questions (rows 9-10 in Table \ref{t:redundant}). Since permutations ranking from $i$ to $j$ answer the same question as the intersection of HDRs corresponding to the same treatments and single consecutive ranks (e.g., (A, B, C)$_i^{j=i+2} \iff$ A$_{\{i\}} \cap$ B$_{\{i+1\}} \cap$ C$_{\{j=i+2\}}$), the collection of HDRs (e.g., A$_{\{i\}}$, B$_{\{i+1\}}$, C$_{\{j=i+2\}}$), which represent their union, are redundant by this ranked permutation (row 9 in Table \ref{t:redundant}). Similarly, a combination ranking from $i$ to $j$ answers the same question as the intersection of HDRs corresponding to the same treatments and set of consecutive ranks (e.g., \{A, B, C\}$_i^{j} \iff$ A$_{\{i,\dots,j\}} \cap$ B$_{\{i,\dots,j\}} \cap$ C$_{\{i,\dots,j\}}$), so the collection of such HDRs is redundant by this ranked combination (row 10 in Table \ref{t:redundant}).

\subsection{Trimming Redundancies through Implicit Partition Blocks}\label{sm:2_1trimpart} \par

A binary treatment hierarchy question $\mathcal{Q}$ involving all $n$ treatments is only expected to be credible at a high threshold (e.g., $\tau = 0.95)$ when there is little to no overlap in the confidence or credible intervals of their relative effects vs. a reference treatment. This does not frequently occur in practice \citep{wigle2025}. However, such ``full" binary treatment hierarchy questions may be partitioned into $k \in \{2, ..., n\}$ blocks of size $\in \{1, 2, ..., n-1\}$ which may be credible. Here, we focus on partition blocks involving ranked hierarchies (i.e., ranked permutations, ranked combinations, and/or HDRs involving one rank), as knowledge of the ordering of blocks within a partition may enable us to infer one block based on $k-1$ blocks. Moreover, if the joint probability of $k-1$ ordered blocks within a partition is credible, then the remaining block is also credible, but it may be considered redundant as it is implied by the $k-1$ blocks. Note that, within a partition, a block consisting of a ranked permutation cannot be inferred by other blocks as the ordering of treatments within the permutation would be unknown. As such, only ranked combinations or HDRs can be redundant since ordering does not matter or is not applicable within these types of treatment hierarchy questions.

We consider three general categories of partitioning that lead to redundancies. The first is one where there are two partition blocks, one involving ranks 1 through $i$, the second involving ranks $i+1$ through $n$ and the empirical probability of these blocks is equivalent. We refer to these as ``top/bottom partition blocks". Within hierarchy question types, this is only possible for partitions consisting of ranked combinations (row 11 in Table \ref{t:redundant}); for clarity in the final output, we only recommend trimming ranked combinations that are made redundant by a redundant ranked combination. Across hierarchy question types, this may occur between a single rank HDR and a ranked combination (rows 12ab in Table \ref{t:redundant}), in which case the single rank HDR would be preferred as it is more concise. Additionally, this may occur between a single rank HDR and a ranked permutation (row 13 in Table \ref{t:redundant}), where the ranked permutation would be prioritized for output as it provides more information. 

The second type of partitioning we consider for redundancy trimming is one involving three partitions blocks: two that encompass the tail ranks (e.g., 1 through $i$ and $j$ through $n$, $1 \leq i < j \leq n$) and one encompassing all other ranks between these tails (e.g., $(i+1)$ through $(j-1)$). We refer to these as ``tail and middle partition blocks". These types of redundancies may occur when a ranked combination forms the middle partition block, while either two single rank HDRs form the tail ranks (row 14 in Table \ref{t:redundant}), a single rank HDR and ranked permutation form the tail ranks (row 15 in Table \ref{t:redundant}), or two ranked permutations form the tail ranks (row 16 in Table \ref{t:redundant}). However, since the empirical probabilities of each block are not necessarily the same, the joint empirical probability of the tail partition blocks needs to be calculated to assess if they are jointly credible and hence imply the credibility of the potentially redundant middle partition block. The joint empirical probability of two tail partition blocks is bounded by the empirical probabilities of their individual blocks outputted by the algorithms described in Section \ref{s:algo_descriptions} of the main manuscript: 
\begin{align}
    \ep(\mathcal{Q}_i \cap \mathcal{Q}_j) \geq min\{\ep(\mathcal{Q}_i), \ep(\mathcal{Q}_j)\} - (1 - max\{\ep(\mathcal{Q}_i), \ep(\mathcal{Q}_j)\}) = \ep(\mathcal{Q}_i) + \ep(\mathcal{Q}_j) - 1.
    \label{e:joint_HQ}
\end{align}
If $\ep(\mathcal{Q}_i) + \ep(\mathcal{Q}_j) - 1 \geq \tau$, then $\ep(\mathcal{Q}_i \cap \mathcal{Q}_j) \geq \tau$. When this threshold is not met, then the joint empirical probability would need to be determined through post-hoc calculations using the matrix of sampled relative effects. Note that because the tail partition blocks may not appear side-by-side in the final output, the implied (potentially trimmed) middle partition block may not be obvious to a researcher. As such, we recommend only trimming middle partition blocks that correspond to redundant (and hence trimmed) tail partition blocks.

In the third category, a partition block may be further divided into more informative partition blocks. This is particularly true for ranked combinations, where the position of the block within the hierarchy is known, but the ordering of treatments within the block is unknown. If a ranked combination can be split into a ranked permutation and smaller ranked combination (row 17 in Table \ref{t:redundant}), or a single rank HDR and smaller ranked combination (row 18 in Table \ref{t:redundant}), which are jointly credible, then it may be considered redundant. We can check joint credibility of split partition blocks using result \eqref{e:joint_HQ}. From the algorithm described in Section \ref{s:algo1} of the main manuscript, we know that the empirical probability of a ranked combination $\mathcal{C}_i^j$ is the sum of the empirical probabilities of ranked permutations $\mathcal{P}_i^j$ involving the same treatments (e.g., $\ep(\{A, B\}_i^j$) = $\ep$((A, B)$_i^j$) + $\ep$((B, A)$_i^j$)) and so $\ep(\mathcal{P}_i^j) \leq \ep(\mathcal{C}_i^j)$. In addition, a size-$s$ ranked combination has an empirical probability less than or equal to that of a bigger ranked combination involving additional treatments and a wider ranking range (e.g., $\ep$(\{C, D\}$_{j-1}^j) \leq \ep$(\{A, B, C, D\}$_i^j)$. As such, in a network involving at least four treatments A, B, C, and D, 
\begin{align*}
    \ep(\{\text{A}, \text{B}, \text{C}, \text{D}\}_i^j) \geq \ep((\text{A}, \text{B})_i^{i+1} \cap \{\text{C}, \text{D}\}_{j-1}^j).
\end{align*}
By result \eqref{e:joint_HQ}, if $\ep$((A, B)$_i^{i+1}$) + $\ep$(\{C, D\}$_{j-1}^j) - 1 \geq \tau$, then $\ep$(\{A, B, C, D\}$_i^j) \geq \tau$. Using similar logic, we can show that 
\begin{align*}
    \ep({\{\text{A}, \text{B}, \text{C}\}_i^j}) \geq \ep(\text{A}_{\{i\}} \cap \{\text{B}, \text{C}\}_{j-1}^j)
\end{align*}
and so by result \eqref{e:joint_HQ}, if $\ep$(A$_{\{i\}}) + \ep$(\{B, C\}$_{j-1}^j) - 1 \geq \tau$, then $\ep$(\{\{A, B, C\}$_i^j\}) \geq \tau$. 

\clearpage

\section{Guidance on Choosing $\tau$ and $K$} \label{sm:3_tau_K_guidance}

A conventional choice for $\tau$ would be 0.95, aligning with Fisher’s historical recommendation \citep{fisher}. For users interested in exploring lower thresholds, we suggest first applying the higher threshold to assess computational feasibility. Among the three algorithmic approaches, the one presented in Section \ref{s:algo2} of the main manuscript is the most sensitive to the choice of $\tau$ in terms of computation time, particularly when there is substantial overlap in the marginal posterior distributions of the relative treatment effects. In contrast, the other two approaches in Sections \ref{s:algo1} and \ref{s:algo3} of the main manuscript are relatively unaffected by changes in $\tau$.

Users must also select the number of samples, $K$, to use. In most practical settings, a high threshold $\tau$ (e.g., 0.95) would be chosen, which means a relatively small $K$ can still yield low Monte Carlo error in the empirical probabilities. Assuming independent samples—as in frequentist NMA—the Monte Carlo standard error in estimating a posterior probability of $\tau$ or above is at most $\sqrt{\tau(1-\tau)/K}$. For instance, with $\tau = 0.95$ and $K$=500 the standard error is below 0.01. However, Bayesian NMA typically yields correlated MCMC samples, in which case $K$ must represent the number of \emph{effective} samples to accurately reflect the Monte Carlo uncertainty \citep{gelman2014}. Alternatively, to gain computational time, the MCMC chains can be run for longer durations and then thinned to yield effectively independent samples, allowing a smaller value of $K$.

Due to sampling variability, some treatment hierarchy questions with true posterior probabilities exceeding the threshold $\tau$ may yield $\ep$s that fall just below $\tau$. To evaluate the robustness of the results and reduce the risk of missing such near-threshold cases, a sensitivity analysis using a slightly lower threshold, denoted $\tau^*$, is recommended. Based on the Central Limit Theorem, we suggest using $\tau^* = \tau -2\sqrt{\tau(1-\tau)/K}$, which ensures that the probability of missing a true positive remains below 2.5\% at each $\ep$ evaluation. This consideration is most important for the algorithm used to uncover partial hierarchies (Section \ref{s:algo2} of the main manuscript), given its iterative reliance on $\ep$ values.

\clearpage

\section{Toy Example: Sample Inputs for Algorithms}\label{sm:4_toyexample}

A sample of the inputs required for each algorithm described in Sections \ref{s:algo1}-\ref{s:algo2} of the main manuscript are provided below, which correspond to a specific toy example involving 5 treatments A, B, C, D, and E.

\subsection{Algorithm 1}\label{sm:4_1algo1}

The matrix of treatment hierarchies, based on $M_2$ in Section \ref{sm:4_2algo2}:

\begin{equation*}
M_1 =
    \begin{bNiceArray}[
  first-row,code-for-first-row=\scriptstyle,
  first-col,code-for-first-col=\scriptstyle,
  ]{ccccc}
         & 1 & 2 & 3 & 4 & 5 \\ 
        1 & \text{A} & \text{B} & \text{C} & \text{D} & \text{E} \\
        2 & \text{A} & \text{B} & \text{C} & \text{D} & \text{E} \\
        3 & \text{A} & \text{B} & \text{C} & \text{D} & \text{E} \\
        4 & \text{A} & \text{B} & \text{D} & \text{C} & \text{E} \\
        5 & \text{A} & \text{B} & \text{C} & \text{D} & \text{E} \\
        6 & \text{A} & \text{B} & \text{D} & \text{E} & \text{C} \\
        \vdots & \vdots & \vdots & \vdots & \vdots & \vdots \\
        1000 & \text{A} & \text{B} & \text{C} & \text{D} & \text{E}
    \end{bNiceArray}
\end{equation*}

\subsection{Algorithm 2} \label{sm:4_2algo2}

The matrix of sampled relative effects vs. a reference treatment, which was selected to be A in the demonstration:

\begin{equation*}
M_2 =
    \begin{bNiceArray}[
  first-row,code-for-first-row=\scriptstyle,
  first-col,code-for-first-col=\scriptstyle,
  ]{ccccc}
         & \text{A} & \text{B} & \text{C} & \text{D} & \text{E} \\ 
        1 & 0 & 0.48 & 0.67 & 1.45 & 1.60 \\
        2 & 0 & 0.16 & 0.67 & 1.18 & 1.61 \\
        3 & 0 & 0.11 & 1.02 & 1.22 & 1.37 \\
        4 & 0 & 0.17 & 1.15 & 1.08 & 1.52 \\
        5 & 0 & 0.29 & 0.73 & 1.01 & 1.50 \\
        6 & 0 & 0.19 & 1.71 & 1.04 & 1.30 \\
        \vdots & \vdots & \vdots & \vdots & \vdots & \vdots \\
        1000 & 0 & 0.18 & 0.88 & 1.26 & 1.28 
    \end{bNiceArray}
\end{equation*}

\clearpage

\section{Toy Example: Demonstration of Algorithm to Determine HDRs} \label{sm:5_algo3}

To determine the HDRs for each treatment, a data frame consisting of the probabilities of each treatment having the $1^\text{st}$ through $n^\text{th}$ rank is required. For the toy example, a sample of the data frame is as follows:

\begin{align*}
    \begin{array}{ccc}
        Treatment & Rank & \ep \\
        \text{A} & 1 & 0.938 \\
        \text{A} & 2 & 0.060 \\
        \vdots & \vdots & \vdots\\
        \text{A} & 5 & 0.000\\
        \text{B} & 1 & 0.056 \\
        \vdots & \vdots & \vdots\\
        \text{B} & 5 & 0.000\\
        \vdots & \vdots & \vdots\\
        \text{E} & 1 & 0.000 \\
        \text{E} & 2 & 0.000 \\
        \vdots & \vdots & \vdots\\
        \text{E} & 5 & 0.855
    \end{array}
\end{align*}

For each treatment, we initially assume that all ranks make up the HDR set (e.g., $\text{A}_{\{1, 2, 3, 4, 5\}}$). The sum of the empirical probabilities excluding the smallest is compared with the credibility threshold (e.g., $min\{\ep(\text{A}_{\{i\}})| i = 1, 2, 3, 4, 5\} = 0.000$, $\ep(\text{A}_{\{1,2,3,4\}}) = 1.000$); if the sum is greater than the credibility threshold (e.g., $\tau = 0.80$), then the rank with the smallest empirical probability is removed from the HDR set. This process is repeated where the rank with the next smallest empirical probability is removed from the HDR set (e.g., $min\{\ep(\text{A}_{\{i\}})| i = 1, 2, 3, 4\} = 0.000$, $\ep(\text{A}_{\{1,2,3\}}) = 1.000$), until the removal of a rank from the HDR results in an empirical probability below the credibility threshold or the HDR is empty (e.g., $\text{A}_{\{1\}})$.

\clearpage

\section{Empirical Illustration: Antihypertensive Drugs for Diabetes} \label{sm:6_diabetes}

Replicating the analysis by \citet{dias2018}, a random effects Bayesian NMA model with a binomial likelihood and a cloglog link was fitted to the data using the \texttt{BUGSnet} R package \citep{bugsnet}. To generate the inputs required by the algorithms proposed in this paper, an initial sample of 50,000 iterations on three MCMC chains (each) was first simulated and discarded as burn-in, and an additional 100,000 samples on each chain were generated and thinned by 100 to reduce the sample size while decreasing autocorrelation between samples. A large number of samples were required to ensure sufficient mixing of chains. This yielded an effective sample size of at least 2908 for each parameter, far superseding the minimum effective sample size of 500 which would ensure the Monte Carlo standard error was less than 0.01 for a credibility threshold of $\tau = 0.95$ (see Section \ref{sm:3_tau_K_guidance}).

Table \ref{tab:diab_red_table} lists all binary treatment hierarchy questions that are credible at the $\tau = 0.95$ threshold.

\begin{table}[hb]
\centering
\noindent\adjustbox{max width=\textwidth, max height=0.45\textheight}{%
    \begin{tabular}{llll}
    \hline
        Type & Treatment Hierarchy Question & $\ep$ & Redundant \\
        \hline
        Ranked Combination & \{ACE inhibitor, ARB, CCB, placebo\}$_1^4$ & 0.9773 & FALSE\\
        & \{beta-blocker, diuretic\}$_5^6$ & 0.9773 & FALSE\\
        \hline
        Combination & \{ACE inhibitor, ARB, CCB, placebo\} & 0.9773 & TRUE\\
        & \{beta-blocker, diuretic\} & 0.9780 & TRUE\\
        \hline
        Partial Hierarchy & ARB $>$ diuretic & 1.0000 & TRUE\\
        at MID = 0& ACE inhibitor $>$ diuretic & 1.0000 & TRUE\\ 
        & placebo $>$ diuretic & 0.9997 & TRUE\\
        & ARB $>$ beta-blocker & 0.9997 & TRUE\\
        & ACE inhibitor $>$ beta-blocker & 0.9997 & TRUE\\
        & ARB $>$ CCB & 0.9963 & TRUE\\
        & CCB $>$ diuretic & 0.9957 & TRUE\\
        & placebo $>$ beta-blocker & 0.9910 & TRUE\\
        & CCB $>$ beta-blocker & 0.9893 & TRUE\\
        & ACE inhibitor $>$ CCB & 0.9817 & TRUE\\
        & ARB $>$ placebo & 0.9797 & TRUE\\
        & ARB $>$ CCB $>$ diuretic & 0.9920 & FALSE\\
        & ARB $>$ CCB $>$ beta-blocker & 0.9857 & FALSE\\
        & ARB $>$ placebo $>$ diuretic & 0.9793 & FALSE\\
        & ACE inhibitor $>$ CCB $>$ diuretic & 0.9773 & FALSE\\
        & ACE inhibitor $>$ CCB $>$ beta-blocker & 0.9710 & FALSE\\
        & ARB $>$ placebo $>$ beta-blocker & 0.9707 & FALSE\\
        \hline
        HDR & ARB$_{\{1-2\}}$ & 0.9790 & FALSE\\
        & ACE inhibitor$_{\{1-3\}}$ & 0.9940 & FALSE\\
        & placebo$_{\{2-4\}}$ & 0.9873 & FALSE\\
        & CCB$_{\{3-4\}}$ & 0.9703 & FALSE\\
        & diuretic$_{\{5-6\}}$ & 0.9957 & TRUE \\
        & beta-blocker$_{\{5-6\}}$ & 0.9817 & TRUE\\
        \hline
    \end{tabular}}
    \caption{All treatment hierarchy questions and high density regions regarding antihypertensive drugs for the prevention of diabetes, which exceed credibility threshold 0.95. Abbreviations: ACE = angiotensin-converting enzyme, ARB = angiotensin receptor blockers, CCB = calcium channel blockers, HDR = high density region, MID = minimally important difference. Notation: $\ep$ = empirical probability.}
\label{tab:diab_red_table}
\end{table}

\clearpage

\section{Empirical Illustration: CBT Interventions for Depression} \label{sm:7_cbtdep}

This illustrative example makes use of an NMA dataset comparing the effectiveness of five cognitive behavioural therapy (CBT) delivery formats for depression and two control conditions (waitlist, care as usual) \citep{dmetar, cuijpers2019}. For illustrative purposes only, a fixed effect Bayesian NMA model with a normal likelihood and an identity link was fitted to the data using the \texttt{BUGSnet} R package \citep{bugsnet}, mimicking the analysis presented in Chapter 12 of Harrer et al. (2021) \citep{harrer2021}, although a random effects Bayesian NMA model fits the data better. The results from a fixed effects model presents an opportunity to examine what treatment hierarchy questions are credible at a high threshold when there is a larger degree of separation between the marginal posterior distributions of the relative treatment effects, compared to the diabetes example. These results should not be used to inform clinical practice.

\begin{figure}[ht]
\begin{center}
\includegraphics[width=0.95\textwidth]{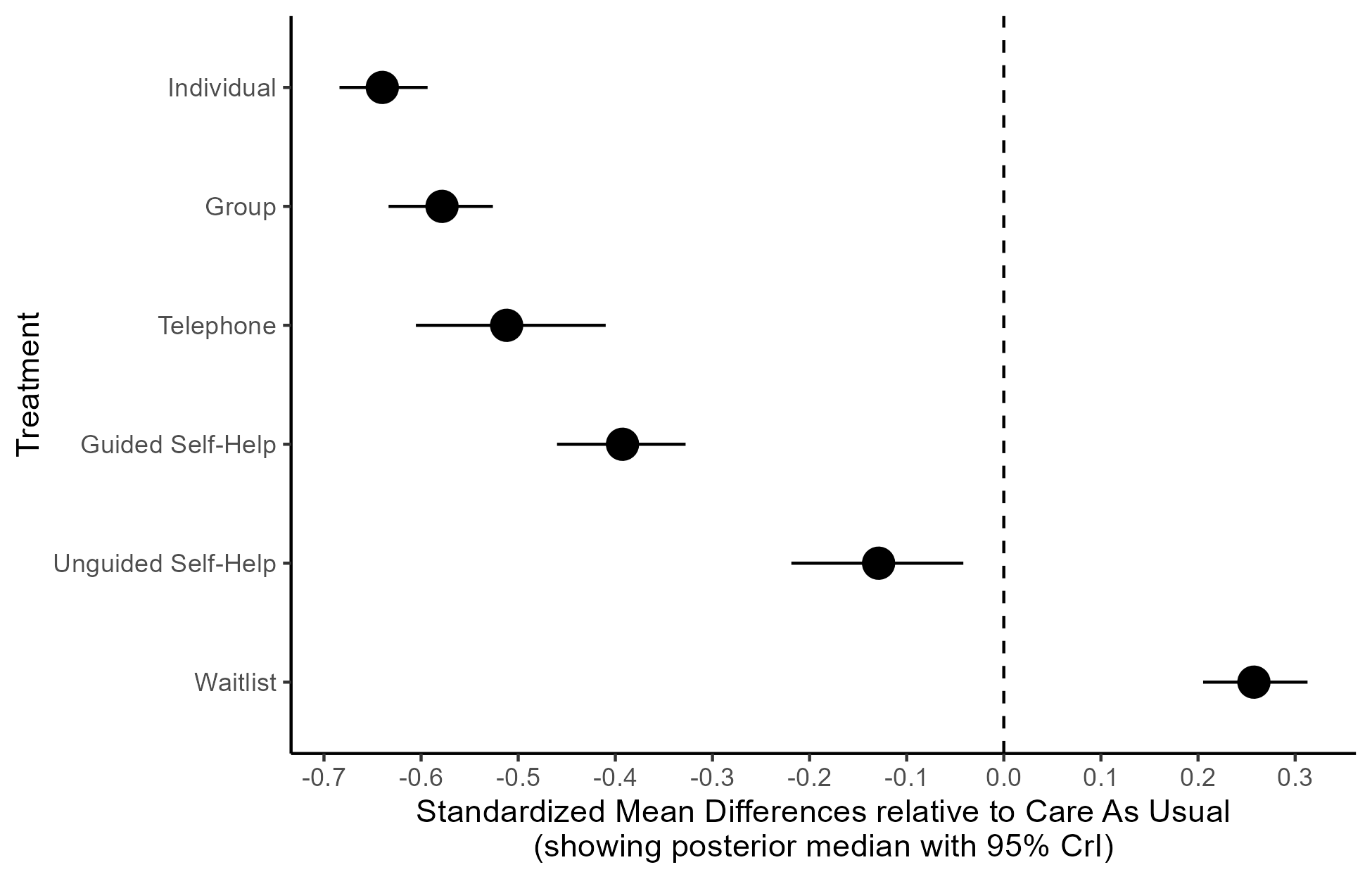}
\caption{Forest plot of the estimated effects of cognitive behavioural therapies vs. care as usual in terms of treating depression in adults based on an NMA \citep{cuijpers2019}. Abbreviations: CrI = credible interval}
\label{fig:cbt_fp}
\end{center}
\end{figure}

Data were inputted and analyzed on the standardized mean difference scale. A forest plot of the estimated relative effects of various CBT formats and waitlist vs. care as usual is presented in Figure \ref{fig:cbt_fp}. Based solely on their point estimates, a hierarchy of all $n=7$ treatments may be $($individual, group, telephone, guided self-help, unguided self-help, care as usual, waitlist$)_1^7$. However, there is some overlap in the credible intervals (CrIs) between individual, group, and telephone, as well as telephone and guided self-help, which adds some uncertainty to this hierarchy, particularly for the first four ranks. We note that the highest degree of overlap in CrIs is between group and telephone CBT.

To generate the inputs required by the algorithms proposed in this paper, an initial sample of 10,000 iterations on three MCMC chains (each) was first simulated and discarded as burn-in, and an additional 20,000 samples on each chain were generated and thinned by 100 to reduce the sample size while decreasing autocorrelation between samples. This yielded an effective sample size of at least 554 for each parameter, thereby ensuring the Monte Carlo standard error was less than 0.01 for a credibility threshold of $\tau = 0.95$ (see Section \ref{sm:3_tau_K_guidance}). Preparing the data for the proposed algorithms took approximately 0.06 seconds. Generating the credible hierarchy question in Sections \ref{s:algo1} - \ref{s:algo3} of the main manuscript took approximately 0.06, 0.19, and 0.01 seconds, respectively, on a laptop with an Intel Core 7 processor and 16.2 GB of RAM.

The algorithms uncovered six ranked permutations, six unranked permutations, 15 ranked combinations, 15 unranked combinations, 88 partial hierarchies, and seven HDRs at a credibility threshold of $\tau = 0.95$ (Table \ref{tab:cbt_red_table}). None of the unranked permutations or combinations added information beyond what was available and credible in the ranked permutations and combinations, and thus the former were redundant. The number of non-redundant hierarchies to consider is much smaller: one ranked permutation, one ranked combination, two partial hierarchies, and one HDRs (Table \ref{tab:cbt_clean_table}). 

In line with the observations drawn from the forest plot (Figure \ref{fig:cbt_fp}), the evidence does not support a complete hierarchy question involving all seven treatments. However, smaller hierarchy questions are credible. It is very likely that guided self-help CBT, unguided self-help CBT, care as usual, and waitlist rank 4th, 5th, 6th, and 7th, respectively. It is also likely that individual CBT ranks 1st. We note that the joint empirical probability of these two hierarchies is at least 0.9833 + 0.9683 - 1 = 0.9516. So, while there was some uncertainty in deciphering the top 4 ranks based on the forest plot, these two hierarchies jointly demonstrate the credibility of individual CBT ranking 1st and guided self-help CBT ranking 4th. In addition, the joint empirical probability of these hierarchies indicate that the remaining treatments, telephone CBT and group CBT, are also likely to be in the second and third ranks, and this can be confirmed by their ranked combination. Nevertheless, it is unclear which rank should be assigned to these two treatments. This is reflected by the partial hierarchies: either is likely to rank worse than individual CBT and better than guided self-help CBT, unguided self help, care as usual, and waitlist. The uncertainty in their specific ranks is due to the overlapping uncertainty in their relative effects vs. care as usual.

A sensitivity analysis using $\tau^* = 0.95 - 2\sqrt{0.95(1-0.95)/500} = 0.9305$ yielded the same credible hierarchy questions before and after trimming, with minimal change in computation time. This is likely due to pattern of overlap between  the marginal posterior distributions of the relative effects. While there was some overlap, it was not enough to pick up additional credible treatment hierarchy questions at the slightly smaller credibility threshold $\tau^*$.

\clearpage

\setlength{\LTcapwidth}{\textwidth}
\begin{longtable}{p{0.2\textwidth}p{0.5\textwidth}p{0.1\textwidth}p{0.1\textwidth}}
    \hline
        Type & Treatment Hierarchy Question & $\ep$ & Redundant\\
        \hline
        Ranked Permutation & (guided self-help, unguided self-help, care as usual, waitlist)$_4^7$ & 0.9833 & FALSE\\
        & (unguided self-help, care as usual, waitlist)$_5^7$ & 1.0000 & TRUE\\
        & (guided self-help, unguided self-help, care as usual)$_4^6$ & 0.9833 & TRUE\\
        & (care as usual, waitlist)$_6^7$ & 1.0000 & TRUE\\ & (unguided self-help, care as usual)$_5^6$ & 1.0000 & TRUE\\
        & (guided self-help, unguided self-help)$_4^5$ & 0.9833 & TRUE\\
        \hline
        Permutation & (guided self-help, unguided self-help, care as usual, waitlist) & 0.9833 & TRUE\\
        & (unguided self-help, care as usual, waitlist) & 1.0000 & TRUE\\
        & (guided self-help, unguided self-help, care as usual) & 0.9833 & TRUE\\
        & (care as usual, waitlist) & 1.0000 & TRUE\\
        & (unguided self-help, care as usual) & 1.0000 & TRUE\\
        & (guided self-help, unguided self-help) & 0.9833 & TRUE\\
        \hline
        Ranked Combination & \{care as usual, group, guided self-help, individual, telephone, unguided self-help\}$_1^6$ & 1.0000 & TRUE\\
        & \{care as usual, group, guided self-help, telephone, unguided self-help, waitlist\}$_2^7$ & 0.9683 & TRUE\\
        & \{group, guided self-help, individual, telephone, unguided self-help\}$_1^5$ & 1.0000 & TRUE\\
        & \{care as usual, group, guided self-help, telephone, unguided self-help\}$_2^6$ & 0.9683 & TRUE\\
        & \{group, guided self-help, individual, telephone\}$_1^4$ & 1.0000 & TRUE\\
        & \{care as usual, guided self-help, unguided self-help, waitlist\}$_4^7$ & 0.9833 & TRUE\\
        & \{group, guided self-help, telephone, unguided self-help\}$_2^5$ & 0.9683 & TRUE\\
        & \{care as usual, unguided self-help, waitlist\}$_5^7$ & 1.0000 & TRUE\\
        & \{care as usual, guided self-help, unguided self-help\}$_4^6$ & 0.9833 & TRUE\\
        & \{group, individual, telephone\}$_1^3$ & 0.9833 & TRUE\\
        & \{group,  self-help, telephone\}$_2^4$ & 0.9683 & TRUE\\
        & \{care as usual, waitlist\}$_6^7$ & 1.0000 & TRUE\\
        & \{care as usual, unguided self-help\}$_5^6$ & 1.0000 & TRUE\\
        & \{guided self-help, unguided self-help\}$_4^5$ & 0.9833 & TRUE\\
        & \{group, telephone\}$_2^3$ & 0.9517 & FALSE\\
        \hline
        Combination & \{care as usual, group, guided self-help, individual, telephone, unguided self-help\} & 1.0000 & TRUE\\
        & \{care as usual, group, guided self-help, telephone, unguided self-help, waitlist\} & 0.9683 & TRUE\\
        & \{group, guided self-help, individual, telephone, unguided self-help\} & 1.0000 & TRUE\\
        & \{care as usual, group, guided self-help, telephone, unguided self-help\} & 0.9683 & TRUE\\
        & \{group, guided self-help, individual, telephone\} & 1.0000 & TRUE\\
        & \{care as usual, guided self-help, unguided self-help, waitlist\} & 0.9833 & TRUE\\
        & \{group, guided self-help, telephone, unguided self-help\} & 0.9683 & TRUE\\
        & \{care as usual, unguided self-help, waitlist\} & 1.0000 & TRUE\\
        & \{group, individual, telephone\} & 0.9833 & TRUE\\
        & \{care as usual, guided self-help, unguided self-help\} & 0.9833 & TRUE\\
        & \{group, guided self-help, telephone\} & 0.9683 & TRUE\\
        & \{care as usual, waitlist\} & 1.0000 & TRUE\\
        & \{care as usual, unguided self-help\} & 1.0000 & TRUE\\
        & \{guided self-help, unguided self-help\} & 0.9833 & TRUE\\
        & \{group, telephone\} & 0.9517 & TRUE\\
        \hline
        Partial Hierarchy & individual $>$ guided self-help & 1.0000 & TRUE\\
        at MID = 0& unguided self-help $>$ care as usual & 1.0000 & TRUE\\
        & telephone $>$ care as usual & 1.0000 & TRUE\\
        & individual $>$ care as usual & 1.0000 & TRUE\\
        & guided self-help $>$ care as usual & 1.0000 & TRUE\\
        & group $>$ care as usual & 1.0000 & TRUE\\
        & unguided self-help $>$ waitlist & 1.0000 & TRUE\\
        & telephone $>$ waitlist & 1.0000 & TRUE\\
        & telephone $>$ unguided self-help & 1.0000 & TRUE\\
        & individual $>$ waitlist & 1.0000 & TRUE\\
        & individual $>$ unguided self-help & 1.0000 & TRUE\\
        & guided self-help $>$ waitlist & 1.0000 & TRUE\\
        & guided self-help $>$ unguided self-help & 1.0000 & TRUE\\
        & group $>$ waitlist & 1.0000 & TRUE\\
        & group $>$ unguided self-help & 1.0000 & TRUE\\
        & group $>$ guided self-help & 1.0000 & TRUE\\
        & care as usual $>$ waitlist & 1.0000 & TRUE\\
        & individual $>$ telephone & 0.9933 & TRUE\\
        & telephone $>$ guided self-help & 0.9833 & TRUE\\
        & individual $>$ group & 0.9750 & TRUE\\
        & individual $>$ guided self-help $>$ waitlist & 1.0000 & TRUE\\
        & individual $>$ guided self-help $>$ unguided self-help & 1.0000 & TRUE\\
        & individual $>$ guided self-help $>$ care as usual & 1.0000 & TRUE\\
        & unguided self-help $>$ care as usual $>$ waitlist & 1.0000 & TRUE\\
        & telephone $>$ care as usual $>$ waitlist & 1.0000 & TRUE\\
        & individual $>$ care as usual $>$ waitlist & 1.0000 & TRUE\\
        & guided self-help $>$ care as usual $>$ waitlist & 1.0000 & TRUE\\
        & group $>$ care as usual $>$ waitlist & 1.0000 & TRUE\\
        & telephone $>$ unguided self-help $>$ waitlist & 1.0000 & TRUE\\
        & telephone $>$ unguided self-help $>$ care as usual & 1.0000 & TRUE\\
        & individual $>$ unguided self-help $>$ waitlist & 1.0000 & TRUE\\
        & individual $>$ unguided self-help $>$ care as usual & 1.0000 & TRUE\\
        & guided self-help $>$ unguided self-help $>$ waitlist & 1.0000 & TRUE\\
        & guided self-help $>$ unguided self-help $>$ care as usual & 1.0000 & TRUE\\
        & group $>$ unguided self-help $>$ waitlist & 1.0000 & TRUE\\
        & group $>$ unguided self-help $>$ care as usual & 1.0000 & TRUE\\
        & group $>$ guided self-help $>$ waitlist & 1.0000 & TRUE\\
        & group $>$ guided self-help $>$ unguided self-help & 1.0000 & TRUE\\
        & group $>$ guided self-help $>$ care as usual & 1.0000 & TRUE\\
        & individual $>$ telephone $>$ waitlist & 0.9933 & TRUE\\
        & individual $>$ telephone $>$ unguided self-help & 0.9933 & TRUE\\
        & individual $>$ telephone $>$ care as usual & 0.9933 & TRUE\\
        & telephone $>$ guided self-help $>$ waitlist & 0.9833 & TRUE\\
        & telephone $>$ guided self-help $>$ unguided self-help & 0.9833 & TRUE\\
        & telephone $>$ guided self-help $>$ care as usual & 0.9833 & TRUE\\
        & individual $>$ telephone $>$ guided self-help & 0.9767 & TRUE\\
        & individual $>$ group $>$ waitlist & 0.9750 & TRUE\\
        & individual $>$ group $>$ unguided self-help & 0.9750 & TRUE\\
        & individual $>$ group $>$ guided self-help & 0.9750 & TRUE\\
        & individual $>$ group $>$ care as usual & 0.9750 & TRUE\\
        & individual $>$ guided self-help $>$ unguided self-help $>$ waitlist & 1.0000 & TRUE\\
        & individual $>$ guided self-help $>$ unguided self-help $>$ care as usual & 1.0000 & TRUE\\
        & individual $>$ guided self-help $>$ care as usual $>$ waitlist & 1.0000 & TRUE\\
        & telephone $>$ unguided self-help $>$ care as usual $>$ waitlist & 1.0000 & TRUE\\
        & individual $>$ unguided self-help $>$ care as usual $>$ waitlist & 1.0000 & TRUE\\
        & guided self-help $>$ unguided self-help $>$ care as usual $>$ waitlist & 1.0000 & TRUE\\
        & group $>$ unguided self-help $>$ care as usual $>$ waitlist & 1.0000 & TRUE\\
        & group $>$ guided self-help $>$ unguided self-help $>$ waitlist & 1.0000 & TRUE\\
        & group $>$ guided self-help $>$ unguided self-help $>$ care as usual & 1.0000 & TRUE\\
        & group $>$ guided self-help $>$ care as usual $>$ waitlist & 1.0000 & TRUE\\
        & individual $>$ telephone $>$ unguided self-help $>$ waitlist & 0.9933 & TRUE\\
        & individual $>$ telephone $>$ unguided self-help $>$ care as usual & 0.9933 & TRUE\\
        & individual $>$ telephone $>$ care as usual $>$ waitlist & 0.9933 & TRUE\\
        & telephone $>$ guided self-help $>$ unguided self-help $>$ waitlist & 0.9833 & TRUE\\
        & telephone $>$ guided self-help $>$ unguided self-help $>$ care as usual & 0.9833 & TRUE\\
        & telephone $>$ guided self-help $>$ care as usual $>$ waitlist & 0.9833 & TRUE\\
        & individual $>$ telephone $>$ guided self-help $>$ waitlist & 0.9767 & TRUE\\
        & individual $>$ telephone $>$ guided self-help $>$ unguided self-help & 0.9767 & TRUE\\
        & individual $>$ telephone $>$ guided self-help $>$ care as usual & 0.9767 & TRUE\\
        & individual $>$ group $>$ unguided self-help $>$ waitlist & 0.9750 & TRUE\\
        & individual $>$ group $>$ unguided self-help $>$ care as usual & 0.9750 & TRUE\\
        & individual $>$ group $>$ guided self-help $>$ waitlist & 0.9750 & TRUE\\
        & individual $>$ group $>$ guided self-help $>$ unguided self-help & 0.9750 & TRUE\\
        & individual $>$ group $>$ guided self-help $>$ care as usual & 0.9750 & TRUE\\
        & individual $>$ group $>$ care as usual $>$ waitlist & 0.9750 & TRUE\\
        & individual $>$ guided self-help $>$ unguided self-help $>$ care as usual $>$ waitlist & 1.0000 & TRUE\\
        & group $>$ guided self-help $>$ unguided self-help $>$ care as usual $>$ waitlist & 1.0000 & TRUE\\
        & individual $>$ telephone $>$ unguided self-help $>$ care as usual $>$ waitlist & 0.9933 & TRUE\\
        & telephone $>$ guided self-help $>$ unguided self-help $>$ care as usual $>$ waitlist & 0.9833 & TRUE\\
        & individual $>$ telephone $>$ guided self-help $>$ unguided self-help $>$ waitlist & 0.9767 & TRUE\\
        & individual $>$ telephone $>$ guided self-help $>$ unguided self-help $>$ care as usual & 0.9767 & TRUE\\
        & individual $>$ telephone $>$ guided self-help $>$ care as usual $>$ waitlist & 0.9767 & TRUE\\
        & individual $>$ group $>$ unguided self-help $>$ care as usual $>$ waitlist & 0.9750 & TRUE\\
        & individual $>$ group $>$ guided self-help $>$ unguided self-help $>$ waitlist & 0.9750 & TRUE\\
        & individual $>$ group $>$ guided self-help $>$ unguided self-help $>$ care as usual & 0.9750 & TRUE\\
        & individual $>$ group $>$ guided self-help $>$ care as usual $>$ waitlist & 0.9750 & TRUE\\
        & individual $>$ telephone $>$ guided self-help $>$ unguided self-help $>$ care as usual $>$ waitlist & 0.9767 & FALSE\\
        & individual $>$ group $>$ guided self-help $>$ unguided self-help $>$ care as usual $>$ waitlist & 0.9750 & FALSE\\
        \hline
        HDRs & individual$_{\{1\}}$ & 0.9683 & FALSE \\
        & telephone$_{\{2-3\}}$ & 0.9767 & TRUE \\
        & group$_{\{2-3\}}$ & 0.9750 & TRUE \\
        & guided self-help$_{\{4\}}$ & 0.9833 & TRUE \\
        & unguided self-help$_{\{5\}}$ & 1.0000 & TRUE \\
        & care as usual$_{\{6\}}$ & 1.0000 & TRUE\\
        & waitlist$_{\{7\}}$ & 1.0000 & TRUE\\
        \hline
    \caption{All treatment hierarchy questions and high density regions regarding cognitive behavioural therapies for depression, which exceed credibility threshold 0.95. Blue bars and asterisks indicate the corresponding rank belongs to the HDR set. Abbreviations: HDR = high density region, MID = minimally important difference. Notation: $\ep$ = empirical probability.}
    \label{tab:cbt_red_table}
\end{longtable}

\begin{table}
\noindent\adjustbox{max width=\textwidth}{%
    \begin{tabular}{lll}
    \hline
        Type & Treatment Hierarchy Question & $\ep$ \\
        \hline
        Ranked Permutation & (guided self-help, unguided self-help, care as usual, waitlist)$_4^7$ & 0.9833 \\
        \hline
        Ranked Combination & \{group, telephone\}$_2^3$ & 0.9517 \\
        \hline
        Partial Hierarchy & individual $>$ telephone $>$ guided self-help $>$ unguided self-help $>$ care as usual $>$ waitlist & 0.9767 \\
        at MID = 0& individual $>$ group $>$ guided self-help $>$ unguided self-help $>$ care as usual $>$ waitlist & 0.9750 \\
        \hline
        HDRs & individual$_{\{1\}}$ & 0.9683 \\
        \hline
    \end{tabular} }
    \caption{Non-redundant hierarchy questions and high density regions regarding antihypertensive drugs for preventing new cases of diabetes, which exceed credibility threshold 0.95. Abbreviations: ACE = angiotensin-converting enzyme, ARB = angiotensin receptor blockers, CCB = calcium channel blockers, HDR = high density region, MID = minimally important difference. Notation: $\ep$ = empirical probability.}
    \label{tab:cbt_clean_table}
\end{table}

\begin{figure}
\begin{center}
\includegraphics[width=1.0\textwidth]{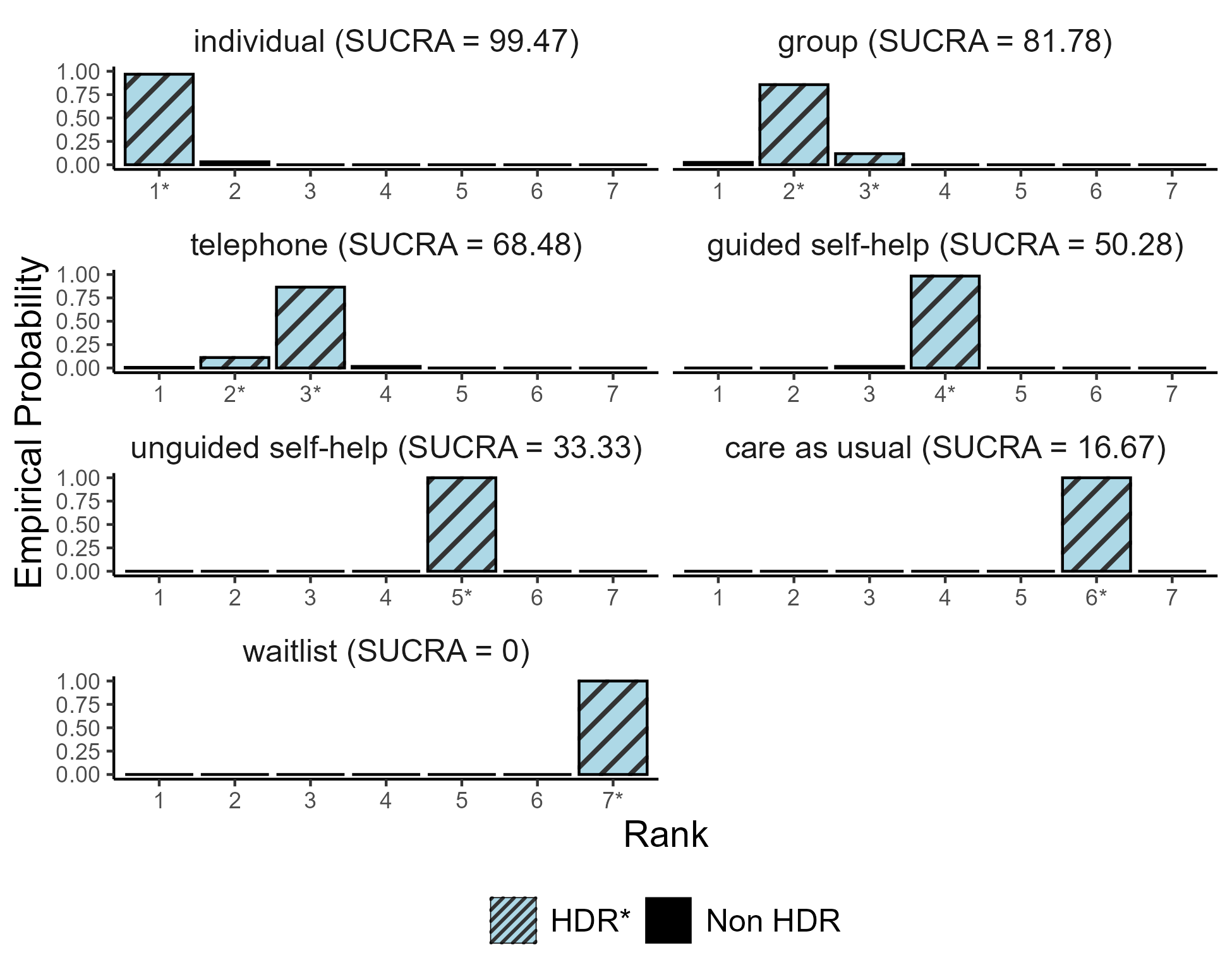}
\caption{Rankograms with HDRs for each treatment in CBT for depression network based on an NMA \citep{cuijpers2019}. Blue bars with black stripes and asterisks indicate the ranks belonging to the HDR set. Abbreviations: ACE = angiotensin-converting enzyme, ARB = angiotensin receptor blockers, CCB = calcium channel blockers, HDR = high density region, SUCRA = surface under the cumulative ranking curve.}
\label{fig:cbt_hdr}
\end{center}
\end{figure}

\clearpage

\end{document}